\documentclass[twocolumn,aps,superscriptaddress,preprintnumbers]{revtex4-2}

\usepackage{amsmath,amssymb}
\usepackage{graphicx}
\usepackage{dcolumn} 
\usepackage{bm} 
\usepackage{soul}
\usepackage{multirow}
\usepackage{setspace}
\usepackage[mathlines]{lineno}
\usepackage{textcomp}
\usepackage{mathcomp}
\usepackage{spverbatim}
\usepackage{float}
\usepackage{wasysym}
\usepackage{array}
\usepackage{subfigure}
\usepackage{physics}
\usepackage{xcolor}

\usepackage{hyperref}
\hypersetup{
    colorlinks=true,       
    linkcolor=blue,        
    citecolor=blue,        
    filecolor=blue,        
    urlcolor=blue,         
    runcolor=blue
}

\usepackage{enumitem,amssymb}

\begin{document}

\title{Lindblad theory for incoherently-driven electron transport in molecular nanojunctions}

\author{Felipe Recabal}

\affiliation{Department of Physics, Universidad de Santiago de Chile, Av. Ecuador 3493, Santiago, Chile.}

\author{Felipe Herrera}

\email{felipe.herrera.u@usach.cl}

\affiliation{Department of Physics, Universidad de Santiago de Chile, Av. Ecuador 3493, Santiago, Chile.}

\affiliation{Millennium Institute for Research in Optics, Concepci\'on, Chile.}

\date{\today}           

\begin{abstract}
We study electron transport in molecular nanojunctions that are driven by incoherent radiation using Markovian quantum dynamics based on the Lindblad quantum master equation. General expressions for the transient electron and photon currents between system and reservoir are derived. For experimentally relevant nanojunction configurations that include on-site Coulomb repulsion, electron tunneling, spontaneous photon emission, and incoherent driving, we show that Lindblad theory can reproduce stationary conductance features reported in the literature such as negative differential conductance, Coulomb blockade, and current-induced light emission. Light-induced currents are predicted for two-site configurations with ground-level tunneling when the incoherent driving rate is comparable with the transfer rate to contact electrodes. Model extensions to include coherent light–matter interaction are suggested.
\end{abstract}

\maketitle

\section{Introduction}

Nanojunctions based on molecules or quantum dots coupled to macroscopic leads are important experimental platforms for studying non-equilibrium electron transport at the nanoscale \cite{tao2010electron,scheer2017molecular,xiang2016molecular,RevModPhys.75.1,gehring2019single,thoss2018perspective,PhysRevResearch.4.043168} . Experimental demonstrations of subtle  effects such as Franck-Condon blockade \cite{leturcq2009franck,burzuri2014franck}, negative differential resistance \cite{xue1999negative,perrin2014large}, quantum interference \cite{guedon2012observation,vazquez2012probing} and few-electron switching \cite{dulic2003one,blum2005molecularly,lara2011light} have stimulated the development of microscopic models that can be used for gaining physical insight and advance the development of next-generation nanoelectronic devices \cite{ALHARBI2015,xiang2016molecular,Liu2020applications,Agarwalla2016}. Since nanojunctions are open systems, quantum master equations are important tools for modeling experimental observables \cite{thoss2018perspective,breuer2002theory,timm2008tunneling}. In addition to describing single-electron open system dynamics \cite{Zeng-Zhao2019,harbola2006quantum,sowa2017environment,sowa2017vibrational}, quantum master equations have been extended to include electron Coulomb interactions \cite{pedersen2005tunneling,esposito2009transport,esposito2010self,li2005quantum,Li2012compensation}, electron-vibration coupling \cite{PhysRevB.83.115414,esposito2009transport,sowa2017vibrational,sowa2017environment}, laser driving \cite{Zhou:2018,Fu2018} and strong electron-photon coupling \cite{Wellnitz2021,orgiu2015conductivity}. 

Quantum master equations in Lindblad form \cite{Manzano2020} preserve the positivity of the system density matrix and allow for simple physical interpretations of dynamical processes in terms of selection rules and decay rates \cite{harbola2006quantum,timm2008tunneling}. However, Lindblad theory is limited in its ability to capture back-action effects such as level shifts and broadenings due to coupling with the contact electrodes, which are better treated using generalized density matrix propagators based on non-equilibrium Green's function \cite{esposito2009transport,esposito2010self,pedersen2005tunneling} and Keldysh contour techniques \cite{timm2008tunneling}. Lindblad quantum master equations can be useful in electron transport systems with competing coherent and incoherent dynamics, which have been shown to deviate from the Landauer formula under certain conditions  \cite{Mejia2022}.


In this work, we study electron transport in single-site and two-site nanojunctions subject to incoherent optical driving, spontaneous emission and Coulomb interaction, using Lindblad quantum master equations. Simple general expressions for the transient and stationary electron current and emitted photon currents are derived in the Born-Markov approximation. For  experimentally relevant cases, we demonstrate that Lindblad theory can capture transport observables reported in the literature, such as negative conductance \cite{perrin2014large} and current-induced light \cite{galperin2005current}. Incoherent optical driving results in light-induced current phenomena \cite{galperin2005current} in agreement experiments \cite{Zhou:2018}, as well as suppression of Coulomb blockade \cite{Fu2018}. 

The rest of the paper is organized as follows: In Sec. \ref{sec:model} we briefly review Lindblad quantum master equations and derive expressions for nanojunction observables. In Sec. \ref{sec:results} we study the behavior of incoherently-driven single-site and two-site systems and validate the model prediction with available experimental data. Conclusions and perspectives for theory extensions are given in Sec. \ref{sec:conclusion}.


\section{Lindblad Transport Model}
\label{sec:model}
We model nanojunctions as open quantum systems with discrete few-electron levels coupled to macroscopic contact electrodes and other bosonic reservoirs (e.g., phonons, photons). Conducting arrays with $N$ sites transport spinless electrons by populating ground and excited levels, $\varepsilon_g$ and $\varepsilon_e$, respectively, at the $i-$th site. The system Hamiltonian $\hat{\mathcal{H}}_S ( \{ \hat{c}_{i,\alpha} \})$ is constructed in terms of fermionic operators $\hat{c}_{i,\alpha}$ which annihilate electrons in state $\alpha= \{g,e\}$ on the $i-$th site. Reservoir-induced transition rates are defined in terms of system eigenstates $\ket{\epsilon}$, satisfying $\hat{\mathcal{H}}_S\ket{\epsilon}=\omega_\epsilon \ket{\epsilon}$ with eigenvalues $\omega_\epsilon$.

The quantum state of the nanojunction is given by the density operator $\hat{\rho}_S$. In the Born-Markov and secular approximations, the state evolution is by the Lindblad quantum master equation \cite{breuer2002theory,Manzano2020} (units of $\hbar=k_{\rm B}=e=1$ are used throughout)
\begin{eqnarray}\label{eq:QMeq}
\frac{{\rm d}}{\rm dt}\hat{\rho}_S  &=&-i[\hat{\mathcal{H}}_S,\hat{\rho}_S]\nonumber \\
&+& \sum_{\epsilon,\epsilon'} \Big\{ f_L(\omega_{\epsilon,\epsilon'})\mathcal{L}[\hat{L}^\dagger_{\Gamma_L}] + \left[ 1-f_L(\omega_{\epsilon,\epsilon'})\right] \mathcal{L}[\hat{L}_{\Gamma_L}]\Big\} \nonumber \\
&+& \sum_{\epsilon,\epsilon'} \Big\{ f_R(\omega_{\epsilon,\epsilon'})\mathcal{L}[\hat{L}^\dagger_{\Gamma_R}] + \left[ 1-f_R(\omega_{\epsilon,\epsilon'})\right] \mathcal{L}[\hat{L}_{\Gamma_R}]\Big\} \nonumber\\
&+& \sum_{\epsilon,\epsilon'} \Big\{ n(\omega_{\epsilon,\epsilon'})\mathcal{L}[\hat{L}^\dagger_{\gamma_{\rm r}}] + \left[1+  n(\omega_{\epsilon,\epsilon'})\right]\mathcal{L}[\hat{L}_{\gamma_{\rm r}}]\Big\} \nonumber\\
&+& \sum_{\epsilon,\epsilon'} \mathcal{L}[\hat{L}^\dagger_{W}],
\end{eqnarray}
where  \mbox{$
\mathcal{L}[\hat{L} ]=\hat{L}\hat{\rho}_S\hat{L}^\dagger-\frac{1}{2}\left( \hat{L}^\dagger\hat{L}\hat{\rho}_S+\hat{\rho}_S\hat{L}^\dagger\hat{L} \right)$} is the general form of the Lindblad dissipator with jump operators $\hat{L}$ whose specific definition is determined by the details of the dissipation channel. The commutator with the system Hamiltonian describes coherent evolution. Equation (\ref{eq:QMeq}) decouples eigenstate populations $\rho_{\epsilon,\epsilon} \equiv \langle \epsilon | \hat{\rho}_S \ket{\epsilon}$ from state coherences $\rho_{\epsilon,\epsilon '} \equiv \langle \epsilon | \hat{\rho}_S\ket{\epsilon'}$ (secular approximation). The rate of coupling to the baths are frequency independent (wide-band approximation), and bath-induced Lamb-shifts of the system energies are neglected.

The second and the third terms in Eq. (\ref{eq:QMeq}) describe incoherent electron transfer between the system and the left and right contact electrodes, respectively. The jump operators for these processes are $\hat L_{\Gamma_{L}}^\dagger \equiv \sqrt{\Gamma_{L}}\bra{\epsilon}( \hat{c}^\dagger_{1,g}+\hat{c}^\dagger_{1,e})\ket{\epsilon'}\ket{\epsilon}\bra{\epsilon'}$ and $\hat L_{\Gamma_{R}}^\dagger \equiv \sqrt{\Gamma_{R}}\bra{\epsilon}( \hat{c}^\dagger_{N,g}+\hat{c}^\dagger_{N,e})\ket{\epsilon'}\ket{\epsilon}\bra{\epsilon'}$. These operators create electrons on the site that is immediately connected to either the left or right contact, respectively, leading to population transfer between system eigenstates. Transition rates depend on the electron transfer rates $\Gamma_L$ and $\Gamma_R$, the chemical potentials $\mu_L$ and $\mu_R$ and the Fermi distribution function $f(\omega )=[ \exp{((\omega-\mu)/T_0)}+1]^{-1}$ evaluated at the system transition frequencies $\omega_{\epsilon,\epsilon'}\equiv \omega_\epsilon-\omega_{\epsilon'}$.  $T_0$ is the effective temperature.

The fourth term in Eq. (\ref{eq:QMeq}) accounts for spontaneous emission of photons due to relaxation of excited electrons, through the jump operator $\hat{L}^\dagger_{\gamma_{\rm r}}=\sqrt{\gamma_{\rm r}}\bra{\epsilon}\sum_{i}\hat{c}_{i,e}^\dagger\hat{c}_{i,g}\ket{\epsilon'}\ket{\epsilon}\bra{\epsilon'}$. The relaxation rate depends on the radiative decay rate $\gamma_{\rm r}$ and Bose photon distribution $n(\omega)=[\exp(\omega/T_0)-1]^{-1}$, evaluated at the system transition frequencies.

The last term of Eq. (\ref{eq:QMeq}) models an incoherent light source that drives electronic transitions from lower to higher energy levels through the jump operator $\hat{L}^\dagger_{W}=\sqrt{W}\bra{\epsilon}\sum_{i}\hat{c}_{i,e}^\dagger\hat{c}_{i,g}\ket{\epsilon'}\ket{\epsilon}\bra{\epsilon'}$, with a driving rate $W$.

When a bias voltage $V_b$ is applied through the nanojunction, the left and right chemical potentials deviate with respect to their Fermi energy $\varepsilon_F$ according to $\mu_L=\varepsilon_F+V_b/2$ and $\mu_R=\varepsilon_F-V_b/2$, respectively. The gradient of chemical potential through the nanojuction generates the net electron currents 
\begin{eqnarray}\label{eq:IL}
I_L(t) &=& -\Gamma_L\sum_{\epsilon,\epsilon'}|\bra{\epsilon}( \hat{c}^\dagger_{1,g}+\hat{c}^\dagger_{1,e})\ket{\epsilon'}|^2\nonumber \\
&\times &\Big\{ [1-f_L(\omega_{\epsilon,\epsilon'})]\rho_{\epsilon,\epsilon}(t) - f_L(\omega_{\epsilon,\epsilon'}) \rho_{\epsilon',\epsilon'} (t)\Big\}, 
\end{eqnarray}
and
\begin{eqnarray}\label{eq:IR}
I_R(t) &=& \Gamma_R\sum_{\epsilon,\epsilon'}|\bra{\epsilon}( \hat{c}^\dagger_{N,g}+\hat{c}^\dagger_{N,e})\ket{\epsilon'}|^2\nonumber \\
&\times & \Big\{ [1-f_R(\omega_{\epsilon,\epsilon'})]\rho_{\epsilon,\epsilon}(t) - f_R(\omega_{\epsilon,\epsilon'}) \rho_{\epsilon',\epsilon'}(t) \Big\},
\end{eqnarray}
in the left and right contacts, respectively. These expressions are derived in Appendix \ref{sec:currents} from the average number of electron in each electrode, with the convention that positive currents move electrons from left to right. Charge conservation requires that the time-dependent left- and right-flowing currents be equal in the steady state, i.e., $I_L(t\rightarrow \infty)=I_R(t\rightarrow \infty)\equiv I$.

Using a similar derivation based on the average number of photons in the radiation field, the photon current due to spontaneous emission of light from the system is given by
\begin{eqnarray}\label{eq:luminescence projected}
j_r(t)&=&\gamma_{\rm r}\sum_{\epsilon,\epsilon'}|\bra{\epsilon}\sum_{i}\hat{c}_{i,e}^\dagger\hat{c}_{i,g}\ket{\epsilon'}|^2\nonumber\\
&&\times \Big\{ [1+n(\omega_{\epsilon,\epsilon'})]\rho_{\epsilon,\epsilon}(t)
 - n(\omega_{\epsilon,\epsilon'}) \rho_{\epsilon',\epsilon'}(t) \Big\}.
\end{eqnarray}

The nanojunctions observables in Eqs. (\ref{eq:IL}), (\ref{eq:IR}) and (\ref{eq:luminescence projected}) depend only on the system populations. Under the secular approximation adopted here, state coherences vanish in the steady state \cite{Zhou:2018,koch2005effects,li2005quantum}. Extensions of the quantum master equation that includes non-secular terms and Lamb-shifts can give different transport results due to contributions from steady-state coherences \cite{Zeng-Zhao2019,harbola2006quantum}.

\section{Results and Discussion} \label{sec:results}
We use Lindblad theory to obtain stationary electron and photon currents for single-site and two-site conducting quantum systems with and without incoherent driving, as a function of bias voltage. Model predictions are compared with available experiments in the literature.

\subsection{Incoherently-driven single-site conductance}
\label{sec:TLS}
\begin{figure*}[t!]
\includegraphics[width=1.0\textwidth]{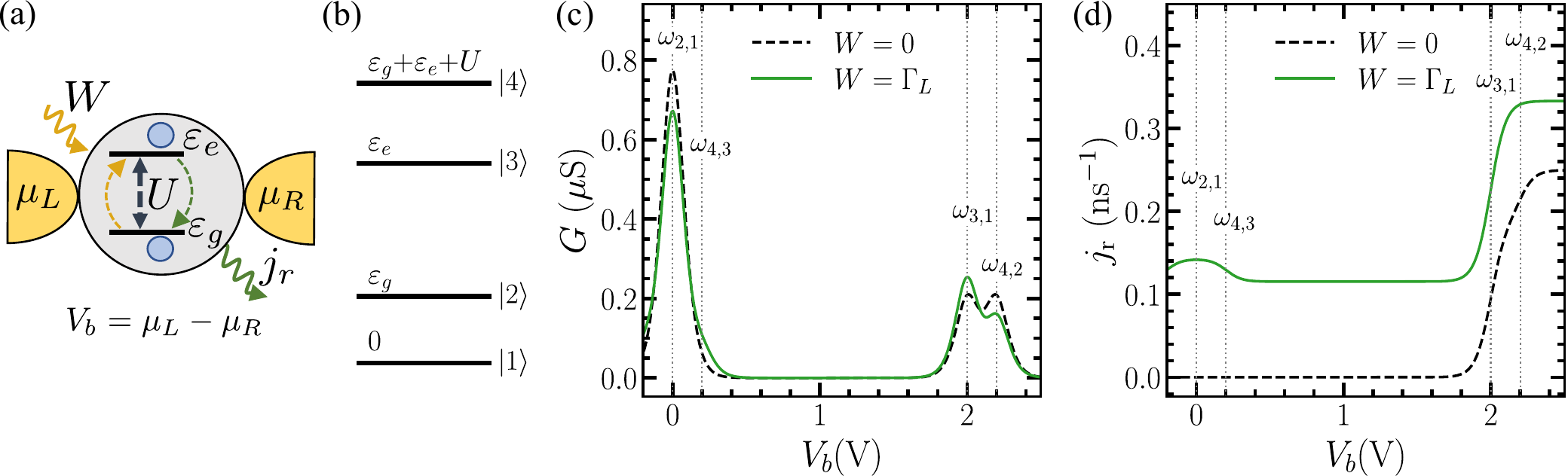}
\caption{{\bf Driven single-site conductance}. (a) Nanojunction scheme with a two-level single site incoherently pumped at rate $W$ and spontaneously emitting photons with flux $j_r$. (b) System energybasis $\{ \ket{\epsilon},\omega_\epsilon \}$. (c) Conductance $G$ as a function of bias voltage $V_b$ for the driven (green line $W=\Gamma_L$) and undriven (black line $W=0$) system. (d) Emitted photon flux $j_{\rm r}$ for the same conditions as in panel (c). Vertical lines represent resonances $\mu_l = \omega_{\epsilon,\epsilon'}$. System parameters are $\{ \varepsilon_F, \varepsilon_g,\varepsilon_e,U \} = \{ 0.5,0.5,1.5,0.1 \}$ eV, $T_0 = 300$ K and $\{ \gamma_r, \Gamma_L, \Gamma_R \} = \{ 1, 10^3,10^3\}$ GHz.}
\label{fig:single_site}
\end{figure*}
Figure \ref{fig:single_site}(a) illustrates a two-level single-site nanojunction subject to incoherent driving at rate $W$ and spontaneous emission at rate $\gamma_r$ (photon flux $j_r$). The system Hamiltonian can be written in terms of fermionic number operators $\hat{n}_{i,\alpha} = \hat{c}_{i,\alpha}^\dagger \hat{c}_{i,\alpha}$ as  
\begin{equation}\label{eq:TLS}
\hat{\mathcal{H}}_S = \varepsilon_g \hat{n}_{1,g} + \varepsilon_e \hat{n}_{1,e} + U \hat{n}_{1,g} \hat{n}_{1,e},
\end{equation}
where $\varepsilon_g$ and $\varepsilon_e$ are the orbital energies and $U$ is the Coulomb interaction energy. Figure \ref{fig:single_site}(b) shows the four-level structure of the system configurations. $\ket{1} = \ket{0_g,0_e}$ ($\omega_1 = 0$) and $\ket{4} = \ket{1_g,1_e}$ ($\omega_4 = \varepsilon_g + \varepsilon_e + U$) represent charged configurations with zero and two electrons, respectively. Eigenstates $\ket{2} = \ket{1_g,0_e}$ ($\omega_2 = \varepsilon_g$) and $\ket{3} = \ket{0_g,1_e}$ ($\omega_3 = \varepsilon_e$) represent neutral configurations of ground and excited electrons. Electron transfer processes with the electrodes result in transitions between eigenstates $\ket{1}\leftrightarrow\ket{2}$ and $\ket{3}\leftrightarrow\ket{4}$ for transfer into the ground orbital, and transitions $\ket{1}\leftrightarrow\ket{3}$ and $\ket{2}\leftrightarrow\ket{4}$ for transfer into the excited orbital. Spontaneous emission and incoherent driving induces transitions $\ket{3}\leftrightarrow \ket{2}$.

Figure \ref{fig:single_site}(c) compares the differential conductance $G = {\rm d} I /{\rm d}V_b$ as a function of $V_b$ for driven and undriven systems. There is a series of conductance peaks when the leads chemical potential matches a transition frequency in the system, giving a resonant voltage of the form $V_b=\pm 2(\omega_{\epsilon,\epsilon'}-\varepsilon_F)$.  The vertical lines mark the resonant voltages for electron transfer involving the ground orbital ($\omega_{2,1}$ and $\omega_{4,3}$) and involving the excited orbital ($\omega_{3,1}$ and $\omega_{4,2}$). The position of the conductance peaks depends on the spectrum of the Hamiltonian and are not affected by the incoherent driving source. However, the amplitude of the conductance peaks are strongly modified when the incoherent driving rate $W$ becomes comparable with the lead-system electron transfer rates $\Gamma_L$ and $\Gamma_R$. These peak shape changes with respect to the undriven case are consequence of the population being driven from the ground to excited neutral configurations.

When electrons flow through the ground orbital, incoherent driving populates the two-electron configuration, increasing the amplitude of the conductance peak at frequency $\omega_{4,3}=\varepsilon_g + U$, while the peak at frequency $\omega_{2,1}=\varepsilon_g$ decreases with respect to an undriven system. When electrons flow through the excited orbital, incoherent driving depopulates two-electron configuration, decreasing the amplitude of the conductance peak at $\omega_{4,2}=\varepsilon_e + U$, and increasing the peak at $\omega_{3,1}=\varepsilon_e$.

 Figure \ref{fig:single_site}(d) shows the corresponding emitted light current $j_{\rm r}$ associated with the conductance curves in panel \ref{fig:single_site}(c). Without incoherent driving, the system spontaneously emits light only when the bias voltage is resonant with the transitions at frequencies $\omega_{3,1}$ and $\omega_{4,2}$, which involve the excited orbital $\varepsilon_e$. This effect corresponds to current-induced light \cite{galperin2005current} and has been observed in Refs.  \cite{qiu2003vibrationally,wu2008intramolecular}. When incoherent driving is present, the emitted light flux increases with respect to the undriven case for all values of $V_b$, because it is constantly moving electrons between ground and excited orbitals.

\begin{figure}[t]
\includegraphics[width=0.46\textwidth]{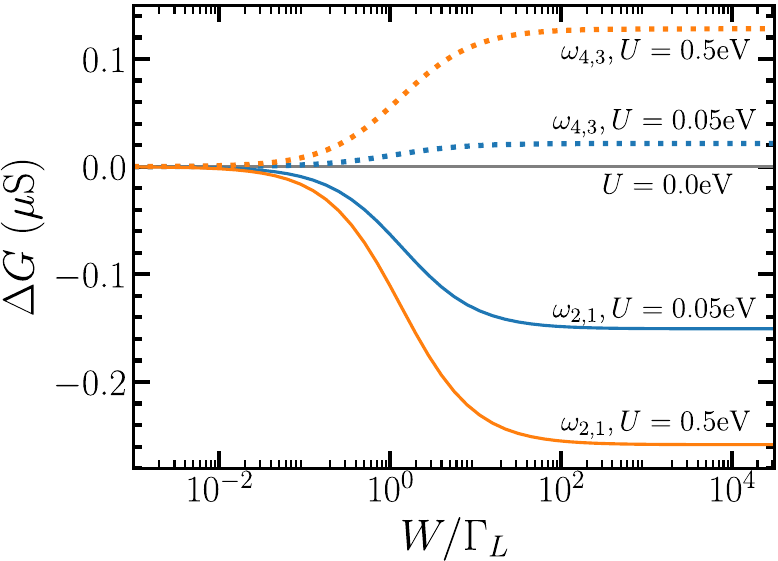}
\caption{{\bf Scaling of driving effect}. Difference of conductance peak $\Delta G$ induced by incoherent driving with respect to the undriven value for the two-level single-site. Conductance peaks associated to transition frequencies $\omega_{2,1}$ (solid lines) and $\omega_{4,3}$ (dashed lines) for different value of Coulomb repulsion $U$. The rest of parameters are the same as in Fig. \ref{fig:single_site}.}
\label{fig:single_site_vs_W}
\end{figure}

The difference of the conductance peak amplitude $\Delta G$ induced by the incoherent driving source, relative to undriven conductance depends primarily on the driving strength $W$. However, for electron transfer involving the ground orbital $\varepsilon_g$, the sensitivity of $\Delta G$ with the magnitude of $W$ depends on the Coulomb interaction strength $U$. For non-interacting electrons ($U=0$), transferring population between ground and excited orbital via incoherent driving does not affect the current $I$ or conductance $G$ at the degenerate frequencies $\omega_{2,1}$ and $\omega_{4,3}$, because electrons are independent. However, when electrons interact ($U>0$), the excited and ground orbital electron transfer channels are no longer independent and the conductivity becomes strongly dependent on $U$. Figure \ref{fig:single_site_vs_W} illustrates this behavior, by showing $\Delta G$ as a function of the relative driving strength $W/\Gamma_L$ for the low bias peaks at $\omega_{2,1}$ and $\omega_{4,3}$, for different values of $U$. For weak driving ($W\ll \Gamma_L$), the conductance is not sensitive to Coulomb interaction. Stronger driving conditions $W\sim \Gamma_L$ are required to modify conductance through the system, corresponding to a light-induced current effect  \cite{galperin2005current}, experimentally observed in \cite{lara2011light,battacharyya2011optical,Zhou:2018}. In Ref. \cite{Zhou:2018} it was shown that Coulomb interaction is required to observe light-induced current. For extreme driving strengths  ($W\gg \Gamma_L$), the magnitude of $\Delta G$ saturates to a value that scales nonlinearly with $U$.

\subsection{Negative conductance in two-site junctions}\label{sec:two-site}
\begin{figure}[t!]
\includegraphics[width=0.44\textwidth]{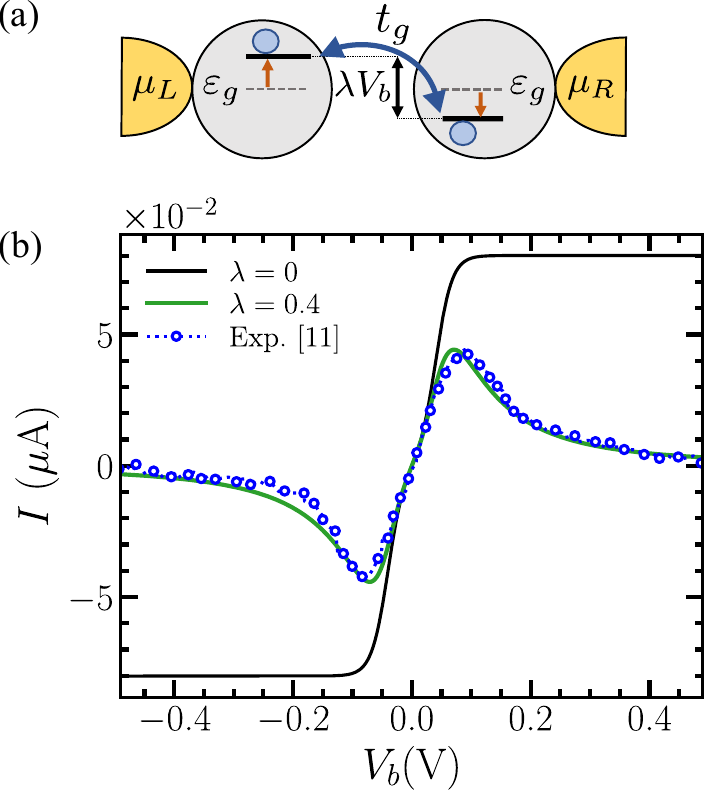}
\caption{{\bf Negative conductance}. (a) Nanojunction scheme of two-sites system with coherent tunneling $t_g$ between the left and the right ground levels, separated in energy by $\lambda V_b$. (b) Current $I$ as a function of the bias voltage $V_b$. Solid curves show Lindblad predictions for $\lambda =0$ (black curve) and $\lambda = 0.4$ (green curve). Experimental results taken from Ref. \cite{perrin2014large} are shown in circles, normalized to $\lambda = 0.4$ theory curve. $t_g=0.02$ eV, $T_0= 80\,{\rm K}$ with other parameters as in Fig. \ref{fig:single_site}.}\label{fig:two_site_NDC}
\end{figure}

Large negative conductance observed in a thiolated arylethynylene molecular junction can be modeled with a two-site conducting array \cite{perrin2014large}, illustrated in Fig. \ref{fig:two_site_NDC}(a). The site ground state orbitals are tuned by a Stark shift proportional to the bias voltage $\lambda V_b$. The system Hamiltonian is given by \cite{perrin2014large}
\begin{eqnarray}\label{eq:two_site}
\hat{\mathcal{H}}_S &=& \left[ \varepsilon_{g} + \lambda\frac{V_b}{2} \right] \hat{n}_{1,g} + \left[ \varepsilon_{g} - \lambda\frac{V_b}{2} \right] \hat{n}_{2,g} \nonumber \\
&+& t_g \left( \hat{c}_{1,g}^\dagger \hat{c}_{2,g}  + \hat{c}_{1,g} \hat{c}_{2,g}^\dagger \right),
\end{eqnarray}
where $\varepsilon_g$ is the ground orbital energy, $\lambda$ is a dimensionless Stark shift parameter, and $t_g$ is the electron tunneling rate. Excited orbitals are not populated in the experiments. The four system eigenstates are composed of charged and neutral states. The zero-electron state $\ket{1}=\ket{0_g,0_g}$ is positively charged, states $\ket{2}$ and $\ket{3}$ are linear combinations of the single electron configurations $\ket{1_g,0_g}$ and $\ket{0_g,1_g}$, and state $\ket{4} = \ket{1_g,1_g}$ is a negatively charged two-electron configuration. To describe negative conductance involving two sites, Coulomb interaction between electrons in different sites is neglected. 

Without Stark shift ($\lambda=0$), the neutral eigenstates are maximally delocalized and given by $\ket{2} = (\ket{1_g,0_g} - \ket{0_g,1_g})/\sqrt{2}$  and $\ket{3} = (\ket{1_g,0_g} + \ket{0_g,1_g})\sqrt{2}$ with energy splitting $2t_g$. This delocalization produces ohmic current at finite bias $V_b$, saturating at high bias as is shown in Fig. \ref{fig:two_site_NDC}(b). Finite Stark shifts change the two-site spectrum, lifting the degeneracy of the configurations $\ket{1_g,0_g}$ and $\ket{0_g,1_g}$  and reducing the degree of charge delocalization. When the energy gap between the states $\lambda V_b$ is much larger than $2t_g$, single-site configurations become fully localized and current is suppressed. Figure \ref{fig:two_site_NDC}(b) shows for $\lambda = 0.4$  that when the bias increases beyond $|V_b|= 2 t_g / \lambda = 0.1\,{\rm V}$, current is suppressed, leading to the negative conductance phenomena reported in Ref. \cite{perrin2014large}. Quantitative agreement with experiments required assuming electron temperature $T_0=80\, {\rm K}$, which is slightly higher than the sample temperature in Ref. \cite{perrin2014large}. 
\begin{figure*}[t]
\includegraphics[width=1.0\textwidth]{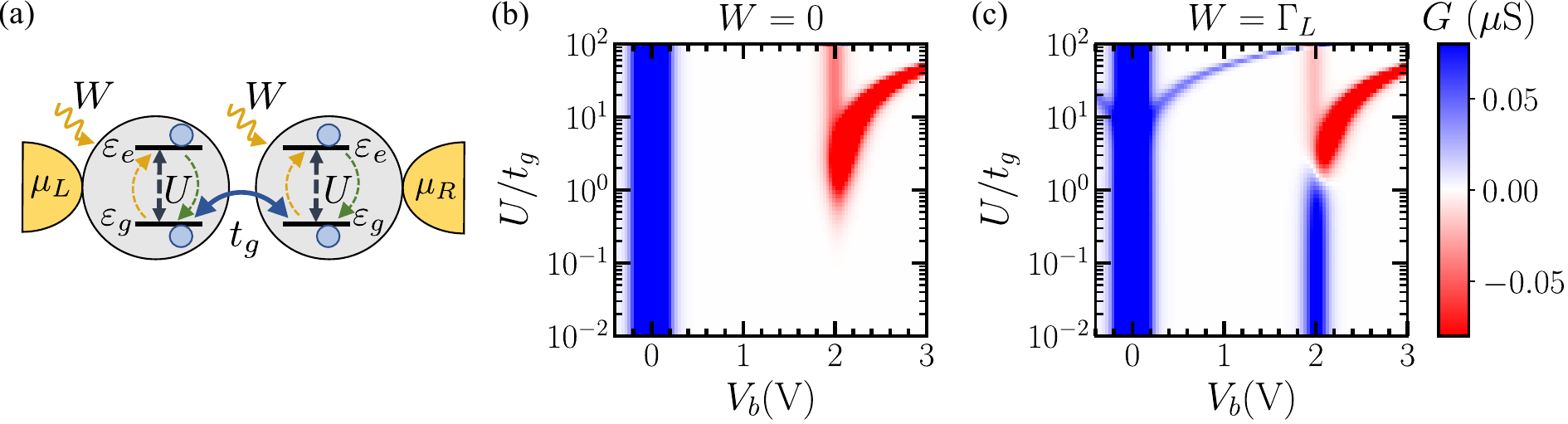}
\caption{{\bf Incoherent driving suppresses Coulomb blockade}. (a) Nanojunction scheme of a driven two-sites system with coherent tunneling rate $t_g$ between the ground orbitals under pumping. Conductance $G$ as a function of bias $V_b$ and Coulomb energy $U$ when the system is (b) undriven and (c) driven at rate $W=\Gamma_L$. Parameters are $t_g=0.01$ eV with other parameters as in Fig. \ref{fig:single_site}.}\label{fig:two_sites_pumping}
\end{figure*}
\subsection{Incoherent driving reduces Coulomb blockade}

Figure \ref{fig:two_sites_pumping}(a) illustrates an incoherently driven two-site two-orbital conducting array with electron tunneling through the ground orbitals. Excited orbitals undergo spontaneous emission to the ground levels and same-site two-electron configurations are subject to Coulomb repulsion. The Hamiltonian for the system can be written as
\begin{eqnarray}\label{eq:H two-site}
\hat{\mathcal{H}}_S &=& \varepsilon_{g} (\hat{n}_{1,g} +  \hat{n}_{2,g}) + \varepsilon_{e}( \hat{n}_{1,e}+ \hat{n}_{2,e})   \\
&+& U  \left( \hat{n}_{1,g} \hat{n}_{1,e} + \hat{n}_{2,g}\hat{n}_{2,e} \right) + t_g \left( \hat{c}_{1,g} \hat{c}_{2,g}^\dagger + \hat{c}_{1,g}^\dagger \hat{c}_{2,g} \right).\nonumber
\end{eqnarray}
Since there is no tunneling in the excited orbitals, electrons are delocalized only between ground orbitals.  For $t_g \ll \{\varepsilon_g,\varepsilon_e\}$, the conductance peaks can be interpreted in terms of transitions between system eigenstates similar to the discussion of the single-site two-level system in Fig. \ref{fig:single_site}. For the two-site problem, we then expect conductance peaks around $V_b\approx  2(\varepsilon_g - \varepsilon_F) $  (0 V) and $V_b \approx 2(\varepsilon_g +U - \varepsilon_F)= 2U$ for electron transfer into the ground levels, and conductance peaks around $V_b \approx  2(\varepsilon_e - \varepsilon_F)$  (2.0 V) and $V_b \approx 2(\varepsilon_e + U - \varepsilon_F)$ (2.0 V+2$U$) for electron transfer into excited levels. These estimates agree with the conductance results in Figs. \ref{fig:two_sites_pumping}(b) and \ref{fig:two_sites_pumping}(c) for the undriven and driven system, respectively.

Figure \ref{fig:two_sites_pumping}(b) shows that for an undriven system, the conductance peak at $V_b\approx 0$ V reflects that electron transport through the ground levels is not affected by Coulomb repulsion because excited levels in each site are unoccupied. The fact that excited states are not populated without driving also explains why despite a low-bias transition around $V_b \approx 2U$ for $U\gg t_g$ is allowed, there is no conductance peak associated with that transition.  At high bias ($V_b \approx  2$ V), a small population of excited electrons is possible through coupling to the leads, which decay to the conducting ground levels via spontaneous emission. Excited electrons are fully localized and generate same-site Coulomb repulsion that competes with coherent tunneling between ground levels, leading Coulomb blockade  when $U\sim t_g$ \cite{PhysRevLett.72.3590,Li2012compensation}. This blockade effect gives a negative conductance peak at $V_b \approx 2{\rm V}$ which splits into a second branch at $V_b \approx 2 {\rm V}+2U $ for $U \gg t_g$. 

Figure \ref{fig:two_sites_pumping}(c) shows that by incoherently driving the nanojunction at rate $W\sim \Gamma_L$, the zero-bias conductance peak is not affected with respect to the undriven case for $U/t_g\ll 1$. Incoherent driving excites ground electrons in both sites, which giving a Coulomb-assisted conductance peak at $V_b \approx 2U$ for $U \gg t_g$. At higher bias ($V_b\approx 2{\rm V}$), incoherent driving gives a positive conductance peak for weakly interacting electrons $U/t_g\ll1 $, because ground state electron transport is supplemented by additional electron transfer channels from the excited orbitals to the leads. The high bias conductance peak is positive only when  $U\ll t_g$. For strongly interacting electrons $U\gtrsim t_g$, Coulomb blockade becomes effective and gives negative conductance  as discussed for the undriven case. This high-bias behaviour is consistent with the results in   Refs. \cite{Fu2018,Li2012compensation} on light-assisted enhancement of electron transport despite Coulomb blockade.  

\section{Conclusions}\label{sec:conclusion}

We developed a simple open quantum system model for electron transport in incoherently-driven nanojunctions based on the Lindblad quantum master equation. We demonstrated good  quantitative agreement of the model with single-molecule experiments in the literature, with sufficient simplicity that allows for transparent physical interpretations. More broadly, using experimentally reasonable parameters, the model is shown to describe reported results on negative conductance, light-induced current, current-induced light and Coulomb blockade.

The relative simplicity of Lindblad quantum master equations facilitate the development of model extensions that include the coupling of conducting electrons to other degrees of freedom such as local vibrations \cite{Liu2020applications}, confined electromagnetic fields in optical cavities \cite{Herrera2016,orgiu2015conductivity,hagenmuller2017cavity,Wellnitz2021,Semenov2019}, as well as strong coherent driving of electrons with laser fields \cite{10.1063/5.0033382}. Recent experiments have reported modifications of the current-voltage characteristics of organic polymers in infrared cavities \cite{kumar2024extraordinary}, opening new possibilities for controlling current-voltage characteristics of materials with cavity quantum electrodynamics effects \cite{herrera2020molecular}. Lindblad modeling of these hybrid light-matter systems could provide physical insights that complement more general non-Markovian approaches based on non-equilibrium Green's functions. 

\begin{acknowledgments}
We thank Leopoldo Mej\'ia for discussions. F.R. is supported by ANID Doctoral Scholarship 21221970 and F.H. by ANID through grants FONDECYT Regular No. 1221420 and the Millennium Science Initiative Program ICN17\_012.
\end{acknowledgments}

\bibliography{electrontransport}

\providecommand{\noopsort}[1]{}\providecommand{\singleletter}[1]{#1}%
\begin{thebibliography}{49}%
\makeatletter
\providecommand \@ifxundefined [1]{%
 \@ifx{#1\undefined}
}%
\providecommand \@ifnum [1]{%
 \ifnum #1\expandafter \@firstoftwo
 \else \expandafter \@secondoftwo
 \fi
}%
\providecommand \@ifx [1]{%
 \ifx #1\expandafter \@firstoftwo
 \else \expandafter \@secondoftwo
 \fi
}%
\providecommand \natexlab [1]{#1}%
\providecommand \enquote  [1]{``#1''}%
\providecommand \bibnamefont  [1]{#1}%
\providecommand \bibfnamefont [1]{#1}%
\providecommand \citenamefont [1]{#1}%
\providecommand \href@noop [0]{\@secondoftwo}%
\providecommand \href [0]{\begingroup \@sanitize@url \@href}%
\providecommand \@href[1]{\@@startlink{#1}\@@href}%
\providecommand \@@href[1]{\endgroup#1\@@endlink}%
\providecommand \@sanitize@url [0]{\catcode `\\12\catcode `\$12\catcode
  `\&12\catcode `\#12\catcode `\^12\catcode `\_12\catcode `\%12\relax}%
\providecommand \@@startlink[1]{}%
\providecommand \@@endlink[0]{}%
\providecommand \url  [0]{\begingroup\@sanitize@url \@url }%
\providecommand \@url [1]{\endgroup\@href {#1}{\urlprefix }}%
\providecommand \urlprefix  [0]{URL }%
\providecommand \Eprint [0]{\href }%
\providecommand \doibase [0]{https://doi.org/}%
\providecommand \selectlanguage [0]{\@gobble}%
\providecommand \bibinfo  [0]{\@secondoftwo}%
\providecommand \bibfield  [0]{\@secondoftwo}%
\providecommand \translation [1]{[#1]}%
\providecommand \BibitemOpen [0]{}%
\providecommand \bibitemStop [0]{}%
\providecommand \bibitemNoStop [0]{.\EOS\space}%
\providecommand \EOS [0]{\spacefactor3000\relax}%
\providecommand \BibitemShut  [1]{\csname bibitem#1\endcsname}%
\let\auto@bib@innerbib\@empty
\bibitem [{\citenamefont {Tao}(2006)}]{tao2010electron}%
  \BibitemOpen
  \bibfield  {author} {\bibinfo {author} {\bibfnamefont {N.~J.}\ \bibnamefont
  {Tao}},\ }\bibfield  {title} {\bibinfo {title} {Electron transport in
  molecular junctions},\ }\href {https://doi.org/10.1038/nnano.2006.130}
  {\bibfield  {journal} {\bibinfo  {journal} {Nature Nanotechnology}\ }\textbf
  {\bibinfo {volume} {1}},\ \bibinfo {pages} {173} (\bibinfo {year}
  {2006})}\BibitemShut {NoStop}%
\bibitem [{\citenamefont {Scheer}\ and\ \citenamefont
  {Cuevas}(2017)}]{scheer2017molecular}%
  \BibitemOpen
  \bibfield  {author} {\bibinfo {author} {\bibfnamefont {E.}~\bibnamefont
  {Scheer}}\ and\ \bibinfo {author} {\bibfnamefont {J.~C.}\ \bibnamefont
  {Cuevas}},\ }\href@noop {} {\emph {\bibinfo {title} {Molecular electronics:
  an introduction to theory and experiment}}},\ Vol.~\bibinfo {volume} {15}\
  (\bibinfo  {publisher} {World Scientific},\ \bibinfo {year}
  {2017})\BibitemShut {NoStop}%
\bibitem [{\citenamefont {Xiang}\ \emph {et~al.}(2016)\citenamefont {Xiang},
  \citenamefont {Wang}, \citenamefont {Jia}, \citenamefont {Lee},\ and\
  \citenamefont {Guo}}]{xiang2016molecular}%
  \BibitemOpen
  \bibfield  {author} {\bibinfo {author} {\bibfnamefont {D.}~\bibnamefont
  {Xiang}}, \bibinfo {author} {\bibfnamefont {X.}~\bibnamefont {Wang}},
  \bibinfo {author} {\bibfnamefont {C.}~\bibnamefont {Jia}}, \bibinfo {author}
  {\bibfnamefont {T.}~\bibnamefont {Lee}},\ and\ \bibinfo {author}
  {\bibfnamefont {X.}~\bibnamefont {Guo}},\ }\bibfield  {title} {\bibinfo
  {title} {Molecular-scale electronics: From concept to function},\ }\href
  {https://doi.org/10.1021/acs.chemrev.5b00680} {\bibfield  {journal} {\bibinfo
   {journal} {Chemical Reviews}\ }\textbf {\bibinfo {volume} {116}},\ \bibinfo
  {pages} {4318} (\bibinfo {year} {2016})},\ \bibinfo {note} {pMID: 26979510},\
  \Eprint {https://arxiv.org/abs/https://doi.org/10.1021/acs.chemrev.5b00680}
  {https://doi.org/10.1021/acs.chemrev.5b00680} \BibitemShut {NoStop}%
\bibitem [{\citenamefont {van~der Wiel}\ \emph {et~al.}(2002)\citenamefont
  {van~der Wiel}, \citenamefont {De~Franceschi}, \citenamefont {Elzerman},
  \citenamefont {Fujisawa}, \citenamefont {Tarucha},\ and\ \citenamefont
  {Kouwenhoven}}]{RevModPhys.75.1}%
  \BibitemOpen
  \bibfield  {author} {\bibinfo {author} {\bibfnamefont {W.~G.}\ \bibnamefont
  {van~der Wiel}}, \bibinfo {author} {\bibfnamefont {S.}~\bibnamefont
  {De~Franceschi}}, \bibinfo {author} {\bibfnamefont {J.~M.}\ \bibnamefont
  {Elzerman}}, \bibinfo {author} {\bibfnamefont {T.}~\bibnamefont {Fujisawa}},
  \bibinfo {author} {\bibfnamefont {S.}~\bibnamefont {Tarucha}},\ and\ \bibinfo
  {author} {\bibfnamefont {L.~P.}\ \bibnamefont {Kouwenhoven}},\ }\bibfield
  {title} {\bibinfo {title} {Electron transport through double quantum dots},\
  }\href {https://doi.org/10.1103/RevModPhys.75.1} {\bibfield  {journal}
  {\bibinfo  {journal} {Rev. Mod. Phys.}\ }\textbf {\bibinfo {volume} {75}},\
  \bibinfo {pages} {1} (\bibinfo {year} {2002})}\BibitemShut {NoStop}%
\bibitem [{\citenamefont {Gehring}\ \emph {et~al.}(2019)\citenamefont
  {Gehring}, \citenamefont {Thijssen},\ and\ \citenamefont {van~der
  Zant}}]{gehring2019single}%
  \BibitemOpen
  \bibfield  {author} {\bibinfo {author} {\bibfnamefont {P.}~\bibnamefont
  {Gehring}}, \bibinfo {author} {\bibfnamefont {J.~M.}\ \bibnamefont
  {Thijssen}},\ and\ \bibinfo {author} {\bibfnamefont {H.~S.~J.}\ \bibnamefont
  {van~der Zant}},\ }\bibfield  {title} {\bibinfo {title} {Single-molecule
  quantum-transport phenomena in break junctions},\ }\href
  {https://doi.org/10.1038/s42254-019-0055-1} {\bibfield  {journal} {\bibinfo
  {journal} {Nature Reviews Physics}\ }\textbf {\bibinfo {volume} {1}},\
  \bibinfo {pages} {381} (\bibinfo {year} {2019})}\BibitemShut {NoStop}%
\bibitem [{\citenamefont {Thoss}\ and\ \citenamefont
  {Evers}(2018)}]{thoss2018perspective}%
  \BibitemOpen
  \bibfield  {author} {\bibinfo {author} {\bibfnamefont {M.}~\bibnamefont
  {Thoss}}\ and\ \bibinfo {author} {\bibfnamefont {F.}~\bibnamefont {Evers}},\
  }\bibfield  {title} {\bibinfo {title} {{Perspective: Theory of quantum
  transport in molecular junctions}},\ }\href
  {https://doi.org/10.1063/1.5003306} {\bibfield  {journal} {\bibinfo
  {journal} {The Journal of Chemical Physics}\ }\textbf {\bibinfo {volume}
  {148}},\ \bibinfo {pages} {030901} (\bibinfo {year} {2018})},\ \Eprint
  {https://arxiv.org/abs/https://pubs.aip.org/aip/jcp/article-pdf/doi/10.1063/1.5003306/13555225/030901\_1\_online.pdf}
  {https://pubs.aip.org/aip/jcp/article-pdf/doi/10.1063/1.5003306/13555225/030901\_1\_online.pdf}
  \BibitemShut {NoStop}%
\bibitem [{\citenamefont {Vigneau}\ \emph {et~al.}(2022)\citenamefont
  {Vigneau}, \citenamefont {Monsel}, \citenamefont {Tabanera}, \citenamefont
  {Aggarwal}, \citenamefont {Bresque}, \citenamefont {Fedele}, \citenamefont
  {Cerisola}, \citenamefont {Briggs}, \citenamefont {Anders}, \citenamefont
  {Parrondo}, \citenamefont {Auff\`eves},\ and\ \citenamefont
  {Ares}}]{PhysRevResearch.4.043168}%
  \BibitemOpen
  \bibfield  {author} {\bibinfo {author} {\bibfnamefont {F.}~\bibnamefont
  {Vigneau}}, \bibinfo {author} {\bibfnamefont {J.}~\bibnamefont {Monsel}},
  \bibinfo {author} {\bibfnamefont {J.}~\bibnamefont {Tabanera}}, \bibinfo
  {author} {\bibfnamefont {K.}~\bibnamefont {Aggarwal}}, \bibinfo {author}
  {\bibfnamefont {L.}~\bibnamefont {Bresque}}, \bibinfo {author} {\bibfnamefont
  {F.}~\bibnamefont {Fedele}}, \bibinfo {author} {\bibfnamefont
  {F.}~\bibnamefont {Cerisola}}, \bibinfo {author} {\bibfnamefont {G.~A.~D.}\
  \bibnamefont {Briggs}}, \bibinfo {author} {\bibfnamefont {J.}~\bibnamefont
  {Anders}}, \bibinfo {author} {\bibfnamefont {J.~M.~R.}\ \bibnamefont
  {Parrondo}}, \bibinfo {author} {\bibfnamefont {A.}~\bibnamefont
  {Auff\`eves}},\ and\ \bibinfo {author} {\bibfnamefont {N.}~\bibnamefont
  {Ares}},\ }\bibfield  {title} {\bibinfo {title} {Ultrastrong coupling between
  electron tunneling and mechanical motion},\ }\href
  {https://doi.org/10.1103/PhysRevResearch.4.043168} {\bibfield  {journal}
  {\bibinfo  {journal} {Phys. Rev. Res.}\ }\textbf {\bibinfo {volume} {4}},\
  \bibinfo {pages} {043168} (\bibinfo {year} {2022})}\BibitemShut {NoStop}%
\bibitem [{\citenamefont {Leturcq}\ \emph {et~al.}(2009)\citenamefont
  {Leturcq}, \citenamefont {Stampfer}, \citenamefont {Inderbitzin},
  \citenamefont {Durrer}, \citenamefont {Hierold}, \citenamefont {Mariani},
  \citenamefont {Schultz}, \citenamefont {von Oppen},\ and\ \citenamefont
  {Ensslin}}]{leturcq2009franck}%
  \BibitemOpen
  \bibfield  {author} {\bibinfo {author} {\bibfnamefont {R.}~\bibnamefont
  {Leturcq}}, \bibinfo {author} {\bibfnamefont {C.}~\bibnamefont {Stampfer}},
  \bibinfo {author} {\bibfnamefont {K.}~\bibnamefont {Inderbitzin}}, \bibinfo
  {author} {\bibfnamefont {L.}~\bibnamefont {Durrer}}, \bibinfo {author}
  {\bibfnamefont {C.}~\bibnamefont {Hierold}}, \bibinfo {author} {\bibfnamefont
  {E.}~\bibnamefont {Mariani}}, \bibinfo {author} {\bibfnamefont {M.~G.}\
  \bibnamefont {Schultz}}, \bibinfo {author} {\bibfnamefont {F.}~\bibnamefont
  {von Oppen}},\ and\ \bibinfo {author} {\bibfnamefont {K.}~\bibnamefont
  {Ensslin}},\ }\bibfield  {title} {\bibinfo {title} {Franck--condon blockade
  in suspended carbon nanotube quantum dots},\ }\href
  {https://doi.org/10.1038/nphys1234} {\bibfield  {journal} {\bibinfo
  {journal} {Nature Physics}\ }\textbf {\bibinfo {volume} {5}},\ \bibinfo
  {pages} {327} (\bibinfo {year} {2009})}\BibitemShut {NoStop}%
\bibitem [{\citenamefont {Burzur{\'\i}}\ \emph {et~al.}(2014)\citenamefont
  {Burzur{\'\i}}, \citenamefont {Yamamoto}, \citenamefont {Warnock},
  \citenamefont {Zhong}, \citenamefont {Park}, \citenamefont {Cornia},\ and\
  \citenamefont {van~der Zant}}]{burzuri2014franck}%
  \BibitemOpen
  \bibfield  {author} {\bibinfo {author} {\bibfnamefont {E.}~\bibnamefont
  {Burzur{\'\i}}}, \bibinfo {author} {\bibfnamefont {Y.}~\bibnamefont
  {Yamamoto}}, \bibinfo {author} {\bibfnamefont {M.}~\bibnamefont {Warnock}},
  \bibinfo {author} {\bibfnamefont {X.}~\bibnamefont {Zhong}}, \bibinfo
  {author} {\bibfnamefont {K.}~\bibnamefont {Park}}, \bibinfo {author}
  {\bibfnamefont {A.}~\bibnamefont {Cornia}},\ and\ \bibinfo {author}
  {\bibfnamefont {H.~S.~J.}\ \bibnamefont {van~der Zant}},\ }\bibfield  {title}
  {\bibinfo {title} {Franck--condon blockade in a single-molecule transistor},\
  }\href {https://doi.org/10.1021/nl500524w} {\bibfield  {journal} {\bibinfo
  {journal} {Nano Letters}\ }\textbf {\bibinfo {volume} {14}},\ \bibinfo
  {pages} {3191} (\bibinfo {year} {2014})},\ \bibinfo {note} {pMID: 24801879},\
  \Eprint {https://arxiv.org/abs/https://doi.org/10.1021/nl500524w}
  {https://doi.org/10.1021/nl500524w} \BibitemShut {NoStop}%
\bibitem [{\citenamefont {Xue}\ \emph {et~al.}(1999)\citenamefont {Xue},
  \citenamefont {Datta}, \citenamefont {Hong}, \citenamefont {Reifenberger},
  \citenamefont {Henderson},\ and\ \citenamefont {Kubiak}}]{xue1999negative}%
  \BibitemOpen
  \bibfield  {author} {\bibinfo {author} {\bibfnamefont {Y.}~\bibnamefont
  {Xue}}, \bibinfo {author} {\bibfnamefont {S.}~\bibnamefont {Datta}}, \bibinfo
  {author} {\bibfnamefont {S.}~\bibnamefont {Hong}}, \bibinfo {author}
  {\bibfnamefont {R.}~\bibnamefont {Reifenberger}}, \bibinfo {author}
  {\bibfnamefont {J.~I.}\ \bibnamefont {Henderson}},\ and\ \bibinfo {author}
  {\bibfnamefont {C.~P.}\ \bibnamefont {Kubiak}},\ }\bibfield  {title}
  {\bibinfo {title} {Negative differential resistance in the scanning-tunneling
  spectroscopy of organic molecules},\ }\href
  {https://doi.org/10.1103/PhysRevB.59.R7852} {\bibfield  {journal} {\bibinfo
  {journal} {Phys. Rev. B}\ }\textbf {\bibinfo {volume} {59}},\ \bibinfo
  {pages} {R7852} (\bibinfo {year} {1999})}\BibitemShut {NoStop}%
\bibitem [{\citenamefont {Perrin}\ \emph {et~al.}(2014)\citenamefont {Perrin},
  \citenamefont {Frisenda}, \citenamefont {Koole}, \citenamefont {Seldenthuis},
  \citenamefont {Gil}, \citenamefont {Valkenier}, \citenamefont {Hummelen},
  \citenamefont {Renaud}, \citenamefont {Grozema}, \citenamefont {Thijssen},
  \citenamefont {Duli{\'c}},\ and\ \citenamefont {van~der
  Zant}}]{perrin2014large}%
  \BibitemOpen
  \bibfield  {author} {\bibinfo {author} {\bibfnamefont {M.~L.}\ \bibnamefont
  {Perrin}}, \bibinfo {author} {\bibfnamefont {R.}~\bibnamefont {Frisenda}},
  \bibinfo {author} {\bibfnamefont {M.}~\bibnamefont {Koole}}, \bibinfo
  {author} {\bibfnamefont {J.~S.}\ \bibnamefont {Seldenthuis}}, \bibinfo
  {author} {\bibfnamefont {J.~A.~C.}\ \bibnamefont {Gil}}, \bibinfo {author}
  {\bibfnamefont {H.}~\bibnamefont {Valkenier}}, \bibinfo {author}
  {\bibfnamefont {J.~C.}\ \bibnamefont {Hummelen}}, \bibinfo {author}
  {\bibfnamefont {N.}~\bibnamefont {Renaud}}, \bibinfo {author} {\bibfnamefont
  {F.~C.}\ \bibnamefont {Grozema}}, \bibinfo {author} {\bibfnamefont {J.~M.}\
  \bibnamefont {Thijssen}}, \bibinfo {author} {\bibfnamefont {D.}~\bibnamefont
  {Duli{\'c}}},\ and\ \bibinfo {author} {\bibfnamefont {H.~S.~J.}\ \bibnamefont
  {van~der Zant}},\ }\bibfield  {title} {\bibinfo {title} {Large negative
  differential conductance in single-molecule break junctions},\ }\href
  {https://doi.org/10.1038/nnano.2014.177} {\bibfield  {journal} {\bibinfo
  {journal} {Nature Nanotechnology}\ }\textbf {\bibinfo {volume} {9}},\
  \bibinfo {pages} {830} (\bibinfo {year} {2014})}\BibitemShut {NoStop}%
\bibitem [{\citenamefont {Gu{\'e}don}\ \emph {et~al.}(2012)\citenamefont
  {Gu{\'e}don}, \citenamefont {Valkenier}, \citenamefont {Markussen},
  \citenamefont {Thygesen}, \citenamefont {Hummelen},\ and\ \citenamefont
  {van~der Molen}}]{guedon2012observation}%
  \BibitemOpen
  \bibfield  {author} {\bibinfo {author} {\bibfnamefont {C.~M.}\ \bibnamefont
  {Gu{\'e}don}}, \bibinfo {author} {\bibfnamefont {H.}~\bibnamefont
  {Valkenier}}, \bibinfo {author} {\bibfnamefont {T.}~\bibnamefont
  {Markussen}}, \bibinfo {author} {\bibfnamefont {K.~S.}\ \bibnamefont
  {Thygesen}}, \bibinfo {author} {\bibfnamefont {J.~C.}\ \bibnamefont
  {Hummelen}},\ and\ \bibinfo {author} {\bibfnamefont {S.~J.}\ \bibnamefont
  {van~der Molen}},\ }\bibfield  {title} {\bibinfo {title} {Observation of
  quantum interference in molecular charge transport},\ }\href
  {https://doi.org/10.1038/nnano.2012.37} {\bibfield  {journal} {\bibinfo
  {journal} {Nature Nanotechnology}\ }\textbf {\bibinfo {volume} {7}},\
  \bibinfo {pages} {305} (\bibinfo {year} {2012})}\BibitemShut {NoStop}%
\bibitem [{\citenamefont {Vazquez}\ \emph {et~al.}(2012)\citenamefont
  {Vazquez}, \citenamefont {Skouta}, \citenamefont {Schneebeli}, \citenamefont
  {Kamenetska}, \citenamefont {Breslow}, \citenamefont {Venkataraman},\ and\
  \citenamefont {Hybertsen}}]{vazquez2012probing}%
  \BibitemOpen
  \bibfield  {author} {\bibinfo {author} {\bibfnamefont {H.}~\bibnamefont
  {Vazquez}}, \bibinfo {author} {\bibfnamefont {R.}~\bibnamefont {Skouta}},
  \bibinfo {author} {\bibfnamefont {S.}~\bibnamefont {Schneebeli}}, \bibinfo
  {author} {\bibfnamefont {M.}~\bibnamefont {Kamenetska}}, \bibinfo {author}
  {\bibfnamefont {R.}~\bibnamefont {Breslow}}, \bibinfo {author} {\bibfnamefont
  {L.}~\bibnamefont {Venkataraman}},\ and\ \bibinfo {author} {\bibfnamefont
  {M.~S.}\ \bibnamefont {Hybertsen}},\ }\bibfield  {title} {\bibinfo {title}
  {Probing the conductance superposition law in single-molecule circuits with
  parallel paths},\ }\href {https://doi.org/10.1038/nnano.2012.147} {\bibfield
  {journal} {\bibinfo  {journal} {Nature Nanotechnology}\ }\textbf {\bibinfo
  {volume} {7}},\ \bibinfo {pages} {663} (\bibinfo {year} {2012})}\BibitemShut
  {NoStop}%
\bibitem [{\citenamefont {Duli\ifmmode~\acute{c}\else \'{c}\fi{}}\ \emph
  {et~al.}(2003)\citenamefont {Duli\ifmmode~\acute{c}\else \'{c}\fi{}},
  \citenamefont {van~der Molen}, \citenamefont {Kudernac}, \citenamefont
  {Jonkman}, \citenamefont {de~Jong}, \citenamefont {Bowden}, \citenamefont
  {van Esch}, \citenamefont {Feringa},\ and\ \citenamefont {van
  Wees}}]{dulic2003one}%
  \BibitemOpen
  \bibfield  {author} {\bibinfo {author} {\bibfnamefont {D.}~\bibnamefont
  {Duli\ifmmode~\acute{c}\else \'{c}\fi{}}}, \bibinfo {author} {\bibfnamefont
  {S.~J.}\ \bibnamefont {van~der Molen}}, \bibinfo {author} {\bibfnamefont
  {T.}~\bibnamefont {Kudernac}}, \bibinfo {author} {\bibfnamefont {H.~T.}\
  \bibnamefont {Jonkman}}, \bibinfo {author} {\bibfnamefont {J.~J.~D.}\
  \bibnamefont {de~Jong}}, \bibinfo {author} {\bibfnamefont {T.~N.}\
  \bibnamefont {Bowden}}, \bibinfo {author} {\bibfnamefont {J.}~\bibnamefont
  {van Esch}}, \bibinfo {author} {\bibfnamefont {B.~L.}\ \bibnamefont
  {Feringa}},\ and\ \bibinfo {author} {\bibfnamefont {B.~J.}\ \bibnamefont {van
  Wees}},\ }\bibfield  {title} {\bibinfo {title} {One-way optoelectronic
  switching of photochromic molecules on gold},\ }\href
  {https://doi.org/10.1103/PhysRevLett.91.207402} {\bibfield  {journal}
  {\bibinfo  {journal} {Phys. Rev. Lett.}\ }\textbf {\bibinfo {volume} {91}},\
  \bibinfo {pages} {207402} (\bibinfo {year} {2003})}\BibitemShut {NoStop}%
\bibitem [{\citenamefont {Blum}\ \emph {et~al.}(2005)\citenamefont {Blum},
  \citenamefont {Kushmerick}, \citenamefont {Long}, \citenamefont {Patterson},
  \citenamefont {Yang}, \citenamefont {Henderson}, \citenamefont {Yao},
  \citenamefont {Tour}, \citenamefont {Shashidhar},\ and\ \citenamefont
  {Ratna}}]{blum2005molecularly}%
  \BibitemOpen
  \bibfield  {author} {\bibinfo {author} {\bibfnamefont {A.~S.}\ \bibnamefont
  {Blum}}, \bibinfo {author} {\bibfnamefont {J.~G.}\ \bibnamefont
  {Kushmerick}}, \bibinfo {author} {\bibfnamefont {D.~P.}\ \bibnamefont
  {Long}}, \bibinfo {author} {\bibfnamefont {C.~H.}\ \bibnamefont {Patterson}},
  \bibinfo {author} {\bibfnamefont {J.~C.}\ \bibnamefont {Yang}}, \bibinfo
  {author} {\bibfnamefont {J.~C.}\ \bibnamefont {Henderson}}, \bibinfo {author}
  {\bibfnamefont {Y.}~\bibnamefont {Yao}}, \bibinfo {author} {\bibfnamefont
  {J.~M.}\ \bibnamefont {Tour}}, \bibinfo {author} {\bibfnamefont
  {R.}~\bibnamefont {Shashidhar}},\ and\ \bibinfo {author} {\bibfnamefont
  {B.~R.}\ \bibnamefont {Ratna}},\ }\bibfield  {title} {\bibinfo {title}
  {Molecularly inherent voltage-controlled conductance switching},\ }\href
  {https://doi.org/10.1038/nmat1309} {\bibfield  {journal} {\bibinfo  {journal}
  {Nature Materials}\ }\textbf {\bibinfo {volume} {4}},\ \bibinfo {pages} {167}
  (\bibinfo {year} {2005})}\BibitemShut {NoStop}%
\bibitem [{\citenamefont {Lara-Avila}\ \emph {et~al.}(2011)\citenamefont
  {Lara-Avila}, \citenamefont {Danilov}, \citenamefont {Kubatkin},
  \citenamefont {Broman}, \citenamefont {Parker},\ and\ \citenamefont
  {Nielsen}}]{lara2011light}%
  \BibitemOpen
  \bibfield  {author} {\bibinfo {author} {\bibfnamefont {S.}~\bibnamefont
  {Lara-Avila}}, \bibinfo {author} {\bibfnamefont {A.~V.}\ \bibnamefont
  {Danilov}}, \bibinfo {author} {\bibfnamefont {S.~E.}\ \bibnamefont
  {Kubatkin}}, \bibinfo {author} {\bibfnamefont {S.~L.}\ \bibnamefont
  {Broman}}, \bibinfo {author} {\bibfnamefont {C.~R.}\ \bibnamefont {Parker}},\
  and\ \bibinfo {author} {\bibfnamefont {M.~B.}\ \bibnamefont {Nielsen}},\
  }\bibfield  {title} {\bibinfo {title} {Light-triggered conductance switching
  in single-molecule dihydroazulene/vinylheptafulvene junctions},\ }\href
  {https://doi.org/10.1021/jp205638b} {\bibfield  {journal} {\bibinfo
  {journal} {The Journal of Physical Chemistry C}\ }\textbf {\bibinfo {volume}
  {115}},\ \bibinfo {pages} {18372} (\bibinfo {year} {2011})},\ \Eprint
  {https://arxiv.org/abs/https://doi.org/10.1021/jp205638b}
  {https://doi.org/10.1021/jp205638b} \BibitemShut {NoStop}%
\bibitem [{\citenamefont {Alharbi}\ and\ \citenamefont
  {Kais}(2015)}]{ALHARBI2015}%
  \BibitemOpen
  \bibfield  {author} {\bibinfo {author} {\bibfnamefont {F.~H.}\ \bibnamefont
  {Alharbi}}\ and\ \bibinfo {author} {\bibfnamefont {S.}~\bibnamefont {Kais}},\
  }\bibfield  {title} {\bibinfo {title} {Theoretical limits of photovoltaics
  efficiency and possible improvements by intuitive approaches learned from
  photosynthesis and quantum coherence},\ }\href
  {https://doi.org/https://doi.org/10.1016/j.rser.2014.11.101} {\bibfield
  {journal} {\bibinfo  {journal} {Renewable and Sustainable Energy Reviews}\
  }\textbf {\bibinfo {volume} {43}},\ \bibinfo {pages} {1073} (\bibinfo {year}
  {2015})}\BibitemShut {NoStop}%
\bibitem [{\citenamefont {Liu}\ and\ \citenamefont
  {Segal}(2020)}]{Liu2020applications}%
  \BibitemOpen
  \bibfield  {author} {\bibinfo {author} {\bibfnamefont {J.}~\bibnamefont
  {Liu}}\ and\ \bibinfo {author} {\bibfnamefont {D.}~\bibnamefont {Segal}},\
  }\bibfield  {title} {\bibinfo {title} {Generalized input-output method to
  quantum transport junctions. ii. applications},\ }\href
  {https://doi.org/10.1103/PhysRevB.101.155407} {\bibfield  {journal} {\bibinfo
   {journal} {Phys. Rev. B}\ }\textbf {\bibinfo {volume} {101}},\ \bibinfo
  {pages} {155407} (\bibinfo {year} {2020})}\BibitemShut {NoStop}%
\bibitem [{\citenamefont {Agarwalla}\ \emph {et~al.}(2016)\citenamefont
  {Agarwalla}, \citenamefont {Kulkarni}, \citenamefont {Mukamel},\ and\
  \citenamefont {Segal}}]{Agarwalla2016}%
  \BibitemOpen
  \bibfield  {author} {\bibinfo {author} {\bibfnamefont {B.~K.}\ \bibnamefont
  {Agarwalla}}, \bibinfo {author} {\bibfnamefont {M.}~\bibnamefont {Kulkarni}},
  \bibinfo {author} {\bibfnamefont {S.}~\bibnamefont {Mukamel}},\ and\ \bibinfo
  {author} {\bibfnamefont {D.}~\bibnamefont {Segal}},\ }\bibfield  {title}
  {\bibinfo {title} {Tunable photonic cavity coupled to a voltage-biased double
  quantum dot system: Diagrammatic nonequilibrium green's function approach},\
  }\href {https://doi.org/10.1103/PhysRevB.94.035434} {\bibfield  {journal}
  {\bibinfo  {journal} {Phys. Rev. B}\ }\textbf {\bibinfo {volume} {94}},\
  \bibinfo {pages} {035434} (\bibinfo {year} {2016})}\BibitemShut {NoStop}%
\bibitem [{\citenamefont {Breuer}\ \emph {et~al.}(2002)\citenamefont {Breuer},
  \citenamefont {Petruccione} \emph {et~al.}}]{breuer2002theory}%
  \BibitemOpen
  \bibfield  {author} {\bibinfo {author} {\bibfnamefont {H.-P.}\ \bibnamefont
  {Breuer}}, \bibinfo {author} {\bibfnamefont {F.}~\bibnamefont {Petruccione}},
  \emph {et~al.},\ }\href@noop {} {\emph {\bibinfo {title} {The theory of open
  quantum systems}}}\ (\bibinfo  {publisher} {Oxford University Press on
  Demand},\ \bibinfo {year} {2002})\BibitemShut {NoStop}%
\bibitem [{\citenamefont {Timm}(2008)}]{timm2008tunneling}%
  \BibitemOpen
  \bibfield  {author} {\bibinfo {author} {\bibfnamefont {C.}~\bibnamefont
  {Timm}},\ }\bibfield  {title} {\bibinfo {title} {Tunneling through molecules
  and quantum dots: Master-equation approaches},\ }\href
  {https://doi.org/10.1103/PhysRevB.77.195416} {\bibfield  {journal} {\bibinfo
  {journal} {Phys. Rev. B}\ }\textbf {\bibinfo {volume} {77}},\ \bibinfo
  {pages} {195416} (\bibinfo {year} {2008})}\BibitemShut {NoStop}%
\bibitem [{\citenamefont {Li}\ and\ \citenamefont
  {Leijnse}(2019)}]{Zeng-Zhao2019}%
  \BibitemOpen
  \bibfield  {author} {\bibinfo {author} {\bibfnamefont {Z.-Z.}\ \bibnamefont
  {Li}}\ and\ \bibinfo {author} {\bibfnamefont {M.}~\bibnamefont {Leijnse}},\
  }\bibfield  {title} {\bibinfo {title} {Quantum interference in transport
  through almost symmetric double quantum dots},\ }\href
  {https://doi.org/10.1103/PhysRevB.99.125406} {\bibfield  {journal} {\bibinfo
  {journal} {Phys. Rev. B}\ }\textbf {\bibinfo {volume} {99}},\ \bibinfo
  {pages} {125406} (\bibinfo {year} {2019})}\BibitemShut {NoStop}%
\bibitem [{\citenamefont {Harbola}\ \emph {et~al.}(2006)\citenamefont
  {Harbola}, \citenamefont {Esposito},\ and\ \citenamefont
  {Mukamel}}]{harbola2006quantum}%
  \BibitemOpen
  \bibfield  {author} {\bibinfo {author} {\bibfnamefont {U.}~\bibnamefont
  {Harbola}}, \bibinfo {author} {\bibfnamefont {M.}~\bibnamefont {Esposito}},\
  and\ \bibinfo {author} {\bibfnamefont {S.}~\bibnamefont {Mukamel}},\
  }\bibfield  {title} {\bibinfo {title} {Quantum master equation for electron
  transport through quantum dots and single molecules},\ }\href
  {https://doi.org/10.1103/PhysRevB.74.235309} {\bibfield  {journal} {\bibinfo
  {journal} {Phys. Rev. B}\ }\textbf {\bibinfo {volume} {74}},\ \bibinfo
  {pages} {235309} (\bibinfo {year} {2006})}\BibitemShut {NoStop}%
\bibitem [{\citenamefont {Sowa}\ \emph
  {et~al.}(2017{\natexlab{a}})\citenamefont {Sowa}, \citenamefont {Mol},
  \citenamefont {Briggs},\ and\ \citenamefont {Gauger}}]{sowa2017environment}%
  \BibitemOpen
  \bibfield  {author} {\bibinfo {author} {\bibfnamefont {J.~K.}\ \bibnamefont
  {Sowa}}, \bibinfo {author} {\bibfnamefont {J.~A.}\ \bibnamefont {Mol}},
  \bibinfo {author} {\bibfnamefont {G.~A.~D.}\ \bibnamefont {Briggs}},\ and\
  \bibinfo {author} {\bibfnamefont {E.~M.}\ \bibnamefont {Gauger}},\ }\bibfield
   {title} {\bibinfo {title} {Environment-assisted quantum transport through
  single-molecule junctions},\ }\href {https://doi.org/10.1039/C7CP06237K}
  {\bibfield  {journal} {\bibinfo  {journal} {Phys. Chem. Chem. Phys.}\
  }\textbf {\bibinfo {volume} {19}},\ \bibinfo {pages} {29534} (\bibinfo {year}
  {2017}{\natexlab{a}})}\BibitemShut {NoStop}%
\bibitem [{\citenamefont {Sowa}\ \emph
  {et~al.}(2017{\natexlab{b}})\citenamefont {Sowa}, \citenamefont {Mol},
  \citenamefont {Briggs},\ and\ \citenamefont {Gauger}}]{sowa2017vibrational}%
  \BibitemOpen
  \bibfield  {author} {\bibinfo {author} {\bibfnamefont {J.~K.}\ \bibnamefont
  {Sowa}}, \bibinfo {author} {\bibfnamefont {J.~A.}\ \bibnamefont {Mol}},
  \bibinfo {author} {\bibfnamefont {G.~A.~D.}\ \bibnamefont {Briggs}},\ and\
  \bibinfo {author} {\bibfnamefont {E.~M.}\ \bibnamefont {Gauger}},\ }\bibfield
   {title} {\bibinfo {title} {Vibrational effects in charge transport through a
  molecular double quantum dot},\ }\href
  {https://doi.org/10.1103/PhysRevB.95.085423} {\bibfield  {journal} {\bibinfo
  {journal} {Phys. Rev. B}\ }\textbf {\bibinfo {volume} {95}},\ \bibinfo
  {pages} {085423} (\bibinfo {year} {2017}{\natexlab{b}})}\BibitemShut
  {NoStop}%
\bibitem [{\citenamefont {Pedersen}\ and\ \citenamefont
  {Wacker}(2005)}]{pedersen2005tunneling}%
  \BibitemOpen
  \bibfield  {author} {\bibinfo {author} {\bibfnamefont {J.~N.}\ \bibnamefont
  {Pedersen}}\ and\ \bibinfo {author} {\bibfnamefont {A.}~\bibnamefont
  {Wacker}},\ }\bibfield  {title} {\bibinfo {title} {Tunneling through
  nanosystems: Combining broadening with many-particle states},\ }\href
  {https://doi.org/10.1103/PhysRevB.72.195330} {\bibfield  {journal} {\bibinfo
  {journal} {Phys. Rev. B}\ }\textbf {\bibinfo {volume} {72}},\ \bibinfo
  {pages} {195330} (\bibinfo {year} {2005})}\BibitemShut {NoStop}%
\bibitem [{\citenamefont {Esposito}\ and\ \citenamefont
  {Galperin}(2009)}]{esposito2009transport}%
  \BibitemOpen
  \bibfield  {author} {\bibinfo {author} {\bibfnamefont {M.}~\bibnamefont
  {Esposito}}\ and\ \bibinfo {author} {\bibfnamefont {M.}~\bibnamefont
  {Galperin}},\ }\bibfield  {title} {\bibinfo {title} {Transport in molecular
  states language: Generalized quantum master equation approach},\ }\href
  {https://doi.org/10.1103/PhysRevB.79.205303} {\bibfield  {journal} {\bibinfo
  {journal} {Phys. Rev. B}\ }\textbf {\bibinfo {volume} {79}},\ \bibinfo
  {pages} {205303} (\bibinfo {year} {2009})}\BibitemShut {NoStop}%
\bibitem [{\citenamefont {Esposito}\ and\ \citenamefont
  {Galperin}(2010)}]{esposito2010self}%
  \BibitemOpen
  \bibfield  {author} {\bibinfo {author} {\bibfnamefont {M.}~\bibnamefont
  {Esposito}}\ and\ \bibinfo {author} {\bibfnamefont {M.}~\bibnamefont
  {Galperin}},\ }\bibfield  {title} {\bibinfo {title} {Self-consistent quantum
  master equation approach to molecular transport},\ }\href
  {https://doi.org/10.1021/jp103369s} {\bibfield  {journal} {\bibinfo
  {journal} {The Journal of Physical Chemistry C}\ }\textbf {\bibinfo {volume}
  {114}},\ \bibinfo {pages} {20362} (\bibinfo {year} {2010})},\ \Eprint
  {https://arxiv.org/abs/https://doi.org/10.1021/jp103369s}
  {https://doi.org/10.1021/jp103369s} \BibitemShut {NoStop}%
\bibitem [{\citenamefont {Li}\ \emph {et~al.}(2005)\citenamefont {Li},
  \citenamefont {Luo}, \citenamefont {Yang}, \citenamefont {Cui},\ and\
  \citenamefont {Yan}}]{li2005quantum}%
  \BibitemOpen
  \bibfield  {author} {\bibinfo {author} {\bibfnamefont {X.-Q.}\ \bibnamefont
  {Li}}, \bibinfo {author} {\bibfnamefont {J.}~\bibnamefont {Luo}}, \bibinfo
  {author} {\bibfnamefont {Y.-G.}\ \bibnamefont {Yang}}, \bibinfo {author}
  {\bibfnamefont {P.}~\bibnamefont {Cui}},\ and\ \bibinfo {author}
  {\bibfnamefont {Y.}~\bibnamefont {Yan}},\ }\bibfield  {title} {\bibinfo
  {title} {Quantum master-equation approach to quantum transport through
  mesoscopic systems},\ }\href {https://doi.org/10.1103/PhysRevB.71.205304}
  {\bibfield  {journal} {\bibinfo  {journal} {Phys. Rev. B}\ }\textbf {\bibinfo
  {volume} {71}},\ \bibinfo {pages} {205304} (\bibinfo {year}
  {2005})}\BibitemShut {NoStop}%
\bibitem [{\citenamefont {Li}\ \emph {et~al.}(2012)\citenamefont {Li},
  \citenamefont {Shishodia}, \citenamefont {Fainberg}, \citenamefont {Apter},
  \citenamefont {Oren}, \citenamefont {Nitzan},\ and\ \citenamefont
  {Ratner}}]{Li2012compensation}%
  \BibitemOpen
  \bibfield  {author} {\bibinfo {author} {\bibfnamefont {G.}~\bibnamefont
  {Li}}, \bibinfo {author} {\bibfnamefont {M.~S.}\ \bibnamefont {Shishodia}},
  \bibinfo {author} {\bibfnamefont {B.~D.}\ \bibnamefont {Fainberg}}, \bibinfo
  {author} {\bibfnamefont {B.}~\bibnamefont {Apter}}, \bibinfo {author}
  {\bibfnamefont {M.}~\bibnamefont {Oren}}, \bibinfo {author} {\bibfnamefont
  {A.}~\bibnamefont {Nitzan}},\ and\ \bibinfo {author} {\bibfnamefont {M.~A.}\
  \bibnamefont {Ratner}},\ }\bibfield  {title} {\bibinfo {title} {Compensation
  of coulomb blocking and energy transfer in the current voltage characteristic
  of molecular conduction junctions},\ }\href
  {https://doi.org/10.1021/nl204130d} {\bibfield  {journal} {\bibinfo
  {journal} {Nano Letters}\ }\textbf {\bibinfo {volume} {12}},\ \bibinfo
  {pages} {2228} (\bibinfo {year} {2012})}\BibitemShut {NoStop}%
\bibitem [{\citenamefont {H\"artle}\ and\ \citenamefont
  {Thoss}(2011)}]{PhysRevB.83.115414}%
  \BibitemOpen
  \bibfield  {author} {\bibinfo {author} {\bibfnamefont {R.}~\bibnamefont
  {H\"artle}}\ and\ \bibinfo {author} {\bibfnamefont {M.}~\bibnamefont
  {Thoss}},\ }\bibfield  {title} {\bibinfo {title} {Resonant electron transport
  in single-molecule junctions: Vibrational excitation, rectification, negative
  differential resistance, and local cooling},\ }\href
  {https://doi.org/10.1103/PhysRevB.83.115414} {\bibfield  {journal} {\bibinfo
  {journal} {Phys. Rev. B}\ }\textbf {\bibinfo {volume} {83}},\ \bibinfo
  {pages} {115414} (\bibinfo {year} {2011})}\BibitemShut {NoStop}%
\bibitem [{\citenamefont {Zhou}\ \emph {et~al.}(2018)\citenamefont {Zhou},
  \citenamefont {Wang}, \citenamefont {Xu},\ and\ \citenamefont
  {Dubi}}]{Zhou:2018}%
  \BibitemOpen
  \bibfield  {author} {\bibinfo {author} {\bibfnamefont {J.}~\bibnamefont
  {Zhou}}, \bibinfo {author} {\bibfnamefont {K.}~\bibnamefont {Wang}}, \bibinfo
  {author} {\bibfnamefont {B.}~\bibnamefont {Xu}},\ and\ \bibinfo {author}
  {\bibfnamefont {Y.}~\bibnamefont {Dubi}},\ }\bibfield  {title} {\bibinfo
  {title} {Photoconductance from exciton binding in molecular junctions},\
  }\href {https://doi.org/10.1021/jacs.7b10479} {\bibfield  {journal} {\bibinfo
   {journal} {Journal of the American Chemical Society}\ }\textbf {\bibinfo
  {volume} {140}},\ \bibinfo {pages} {70} (\bibinfo {year} {2018})}\BibitemShut
  {NoStop}%
\bibitem [{\citenamefont {Fu}\ \emph {et~al.}(2018)\citenamefont {Fu},
  \citenamefont {Mosquera}, \citenamefont {Schatz}, \citenamefont {Ratner},\
  and\ \citenamefont {Hsu}}]{Fu2018}%
  \BibitemOpen
  \bibfield  {author} {\bibinfo {author} {\bibfnamefont {B.}~\bibnamefont
  {Fu}}, \bibinfo {author} {\bibfnamefont {M.~A.}\ \bibnamefont {Mosquera}},
  \bibinfo {author} {\bibfnamefont {G.~C.}\ \bibnamefont {Schatz}}, \bibinfo
  {author} {\bibfnamefont {M.~A.}\ \bibnamefont {Ratner}},\ and\ \bibinfo
  {author} {\bibfnamefont {L.-Y.}\ \bibnamefont {Hsu}},\ }\bibfield  {title}
  {\bibinfo {title} {Photoinduced anomalous coulomb blockade and the role of
  triplet states in electron transport through an irradiated molecular
  transistor},\ }\href {https://doi.org/10.1021/acs.nanolett.8b01838}
  {\bibfield  {journal} {\bibinfo  {journal} {Nano Letters}\ }\textbf {\bibinfo
  {volume} {18}},\ \bibinfo {pages} {5015} (\bibinfo {year}
  {2018})}\BibitemShut {NoStop}%
\bibitem [{\citenamefont {Wellnitz}\ \emph {et~al.}(2021)\citenamefont
  {Wellnitz}, \citenamefont {Pupillo},\ and\ \citenamefont
  {Schachenmayer}}]{Wellnitz2021}%
  \BibitemOpen
  \bibfield  {author} {\bibinfo {author} {\bibfnamefont {D.}~\bibnamefont
  {Wellnitz}}, \bibinfo {author} {\bibfnamefont {G.}~\bibnamefont {Pupillo}},\
  and\ \bibinfo {author} {\bibfnamefont {J.}~\bibnamefont {Schachenmayer}},\
  }\bibfield  {title} {\bibinfo {title} {{A quantum optics approach to
  photoinduced electron transfer in cavities}},\ }\href
  {https://doi.org/10.1063/5.0037412} {\bibfield  {journal} {\bibinfo
  {journal} {The Journal of Chemical Physics}\ }\textbf {\bibinfo {volume}
  {154}},\ \bibinfo {pages} {054104} (\bibinfo {year} {2021})},\ \Eprint
  {https://arxiv.org/abs/https://pubs.aip.org/aip/jcp/article-pdf/doi/10.1063/5.0037412/14763012/054104\_1\_online.pdf}
  {https://pubs.aip.org/aip/jcp/article-pdf/doi/10.1063/5.0037412/14763012/054104\_1\_online.pdf}
  \BibitemShut {NoStop}%
\bibitem [{\citenamefont {Orgiu}\ \emph {et~al.}(2015)\citenamefont {Orgiu},
  \citenamefont {George}, \citenamefont {Hutchison}, \citenamefont {Devaux},
  \citenamefont {Dayen}, \citenamefont {Doudin}, \citenamefont {Stellacci},
  \citenamefont {Genet}, \citenamefont {Schachenmayer}, \citenamefont {Genes},
  \citenamefont {Pupillo}, \citenamefont {Samor{\`\i}},\ and\ \citenamefont
  {Ebbesen}}]{orgiu2015conductivity}%
  \BibitemOpen
  \bibfield  {author} {\bibinfo {author} {\bibfnamefont {E.}~\bibnamefont
  {Orgiu}}, \bibinfo {author} {\bibfnamefont {J.}~\bibnamefont {George}},
  \bibinfo {author} {\bibfnamefont {J.~A.}\ \bibnamefont {Hutchison}}, \bibinfo
  {author} {\bibfnamefont {E.}~\bibnamefont {Devaux}}, \bibinfo {author}
  {\bibfnamefont {J.~F.}\ \bibnamefont {Dayen}}, \bibinfo {author}
  {\bibfnamefont {B.}~\bibnamefont {Doudin}}, \bibinfo {author} {\bibfnamefont
  {F.}~\bibnamefont {Stellacci}}, \bibinfo {author} {\bibfnamefont
  {C.}~\bibnamefont {Genet}}, \bibinfo {author} {\bibfnamefont
  {J.}~\bibnamefont {Schachenmayer}}, \bibinfo {author} {\bibfnamefont
  {C.}~\bibnamefont {Genes}}, \bibinfo {author} {\bibfnamefont
  {G.}~\bibnamefont {Pupillo}}, \bibinfo {author} {\bibfnamefont
  {P.}~\bibnamefont {Samor{\`\i}}},\ and\ \bibinfo {author} {\bibfnamefont
  {T.~W.}\ \bibnamefont {Ebbesen}},\ }\bibfield  {title} {\bibinfo {title}
  {Conductivity in organic semiconductors hybridized with the vacuum field},\
  }\href {https://doi.org/10.1038/nmat4392} {\bibfield  {journal} {\bibinfo
  {journal} {Nature Materials}\ }\textbf {\bibinfo {volume} {14}},\ \bibinfo
  {pages} {1123} (\bibinfo {year} {2015})}\BibitemShut {NoStop}%
\bibitem [{\citenamefont {Manzano}(2020)}]{Manzano2020}%
  \BibitemOpen
  \bibfield  {author} {\bibinfo {author} {\bibfnamefont {D.}~\bibnamefont
  {Manzano}},\ }\bibfield  {title} {\bibinfo {title} {A short introduction to
  the lindblad master equation},\ }\href {https://doi.org/10.1063/1.5115323}
  {\bibfield  {journal} {\bibinfo  {journal} {AIP Advances}\ }\textbf {\bibinfo
  {volume} {10}},\ \bibinfo {pages} {025106} (\bibinfo {year}
  {2020})}\BibitemShut {NoStop}%
\bibitem [{\citenamefont {Mej{\'\i}a}\ \emph {et~al.}(2022)\citenamefont
  {Mej{\'\i}a}, \citenamefont {Kleinekath{\"o}fer},\ and\ \citenamefont
  {Franco}}]{Mejia2022}%
  \BibitemOpen
  \bibfield  {author} {\bibinfo {author} {\bibfnamefont {L.}~\bibnamefont
  {Mej{\'\i}a}}, \bibinfo {author} {\bibfnamefont {U.}~\bibnamefont
  {Kleinekath{\"o}fer}},\ and\ \bibinfo {author} {\bibfnamefont
  {I.}~\bibnamefont {Franco}},\ }\bibfield  {title} {\bibinfo {title} {Coherent
  and incoherent contributions to molecular electron transport},\ }\href
  {https://doi.org/10.1063/5.0079708} {\bibfield  {journal} {\bibinfo
  {journal} {The Journal of Chemical Physics}\ }\textbf {\bibinfo {volume}
  {156}},\ \bibinfo {pages} {094302} (\bibinfo {year} {2022})}\BibitemShut
  {NoStop}%
\bibitem [{\citenamefont {Galperin}\ and\ \citenamefont
  {Nitzan}(2005)}]{galperin2005current}%
  \BibitemOpen
  \bibfield  {author} {\bibinfo {author} {\bibfnamefont {M.}~\bibnamefont
  {Galperin}}\ and\ \bibinfo {author} {\bibfnamefont {A.}~\bibnamefont
  {Nitzan}},\ }\bibfield  {title} {\bibinfo {title} {Current-induced light
  emission and light-induced current in molecular-tunneling junctions},\ }\href
  {https://doi.org/10.1103/PhysRevLett.95.206802} {\bibfield  {journal}
  {\bibinfo  {journal} {Phys. Rev. Lett.}\ }\textbf {\bibinfo {volume} {95}},\
  \bibinfo {pages} {206802} (\bibinfo {year} {2005})}\BibitemShut {NoStop}%
\bibitem [{\citenamefont {Koch}\ and\ \citenamefont {von
  Oppen}(2005)}]{koch2005effects}%
  \BibitemOpen
  \bibfield  {author} {\bibinfo {author} {\bibfnamefont {J.}~\bibnamefont
  {Koch}}\ and\ \bibinfo {author} {\bibfnamefont {F.}~\bibnamefont {von
  Oppen}},\ }\bibfield  {title} {\bibinfo {title} {Effects of charge-dependent
  vibrational frequencies and anharmonicities in transport through molecules},\
  }\href {https://doi.org/10.1103/PhysRevB.72.113308} {\bibfield  {journal}
  {\bibinfo  {journal} {Phys. Rev. B}\ }\textbf {\bibinfo {volume} {72}},\
  \bibinfo {pages} {113308} (\bibinfo {year} {2005})}\BibitemShut {NoStop}%
\bibitem [{\citenamefont {Qiu}\ \emph {et~al.}(2003)\citenamefont {Qiu},
  \citenamefont {Nazin},\ and\ \citenamefont {Ho}}]{qiu2003vibrationally}%
  \BibitemOpen
  \bibfield  {author} {\bibinfo {author} {\bibfnamefont {X.~H.}\ \bibnamefont
  {Qiu}}, \bibinfo {author} {\bibfnamefont {G.~V.}\ \bibnamefont {Nazin}},\
  and\ \bibinfo {author} {\bibfnamefont {W.}~\bibnamefont {Ho}},\ }\bibfield
  {title} {\bibinfo {title} {Vibrationally resolved fluorescence excited with
  submolecular precision},\ }\href {https://doi.org/10.1126/science.1078675}
  {\bibfield  {journal} {\bibinfo  {journal} {Science}\ }\textbf {\bibinfo
  {volume} {299}},\ \bibinfo {pages} {542} (\bibinfo {year} {2003})},\ \Eprint
  {https://arxiv.org/abs/https://www.science.org/doi/pdf/10.1126/science.1078675}
  {https://www.science.org/doi/pdf/10.1126/science.1078675} \BibitemShut
  {NoStop}%
\bibitem [{\citenamefont {Wu}\ \emph {et~al.}(2008)\citenamefont {Wu},
  \citenamefont {Nazin},\ and\ \citenamefont {Ho}}]{wu2008intramolecular}%
  \BibitemOpen
  \bibfield  {author} {\bibinfo {author} {\bibfnamefont {S.~W.}\ \bibnamefont
  {Wu}}, \bibinfo {author} {\bibfnamefont {G.~V.}\ \bibnamefont {Nazin}},\ and\
  \bibinfo {author} {\bibfnamefont {W.}~\bibnamefont {Ho}},\ }\bibfield
  {title} {\bibinfo {title} {Intramolecular photon emission from a single
  molecule in a scanning tunneling microscope},\ }\href
  {https://doi.org/10.1103/PhysRevB.77.205430} {\bibfield  {journal} {\bibinfo
  {journal} {Phys. Rev. B}\ }\textbf {\bibinfo {volume} {77}},\ \bibinfo
  {pages} {205430} (\bibinfo {year} {2008})}\BibitemShut {NoStop}%
\bibitem [{\citenamefont {Battacharyya}\ \emph {et~al.}(2011)\citenamefont
  {Battacharyya}, \citenamefont {Kibel}, \citenamefont {Kodis}, \citenamefont
  {Liddell}, \citenamefont {Gervaldo}, \citenamefont {Gust},\ and\
  \citenamefont {Lindsay}}]{battacharyya2011optical}%
  \BibitemOpen
  \bibfield  {author} {\bibinfo {author} {\bibfnamefont {S.}~\bibnamefont
  {Battacharyya}}, \bibinfo {author} {\bibfnamefont {A.}~\bibnamefont {Kibel}},
  \bibinfo {author} {\bibfnamefont {G.}~\bibnamefont {Kodis}}, \bibinfo
  {author} {\bibfnamefont {P.~A.}\ \bibnamefont {Liddell}}, \bibinfo {author}
  {\bibfnamefont {M.}~\bibnamefont {Gervaldo}}, \bibinfo {author}
  {\bibfnamefont {D.}~\bibnamefont {Gust}},\ and\ \bibinfo {author}
  {\bibfnamefont {S.}~\bibnamefont {Lindsay}},\ }\bibfield  {title} {\bibinfo
  {title} {Optical modulation of molecular conductance},\ }\href
  {https://doi.org/10.1021/nl200977c} {\bibfield  {journal} {\bibinfo
  {journal} {Nano Letters}\ }\textbf {\bibinfo {volume} {11}},\ \bibinfo
  {pages} {2709} (\bibinfo {year} {2011})},\ \bibinfo {note} {pMID: 21657259},\
  \Eprint {https://arxiv.org/abs/https://doi.org/10.1021/nl200977c}
  {https://doi.org/10.1021/nl200977c} \BibitemShut {NoStop}%
\bibitem [{\citenamefont {Stafford}\ and\ \citenamefont
  {Das~Sarma}(1994)}]{PhysRevLett.72.3590}%
  \BibitemOpen
  \bibfield  {author} {\bibinfo {author} {\bibfnamefont {C.~A.}\ \bibnamefont
  {Stafford}}\ and\ \bibinfo {author} {\bibfnamefont {S.}~\bibnamefont
  {Das~Sarma}},\ }\bibfield  {title} {\bibinfo {title} {Collective coulomb
  blockade in an array of quantum dots: A mott-hubbard approach},\ }\href
  {https://doi.org/10.1103/PhysRevLett.72.3590} {\bibfield  {journal} {\bibinfo
   {journal} {Phys. Rev. Lett.}\ }\textbf {\bibinfo {volume} {72}},\ \bibinfo
  {pages} {3590} (\bibinfo {year} {1994})}\BibitemShut {NoStop}%
\bibitem [{\citenamefont {Herrera}\ and\ \citenamefont
  {Spano}(2016)}]{Herrera2016}%
  \BibitemOpen
  \bibfield  {author} {\bibinfo {author} {\bibfnamefont {F.}~\bibnamefont
  {Herrera}}\ and\ \bibinfo {author} {\bibfnamefont {F.~C.}\ \bibnamefont
  {Spano}},\ }\bibfield  {title} {\bibinfo {title} {Cavity-controlled chemistry
  in molecular ensembles},\ }\href
  {https://doi.org/10.1103/PhysRevLett.116.238301} {\bibfield  {journal}
  {\bibinfo  {journal} {Phys. Rev. Lett.}\ }\textbf {\bibinfo {volume} {116}},\
  \bibinfo {pages} {238301} (\bibinfo {year} {2016})}\BibitemShut {NoStop}%
\bibitem [{\citenamefont {Hagenm\"uller}\ \emph {et~al.}(2017)\citenamefont
  {Hagenm\"uller}, \citenamefont {Schachenmayer}, \citenamefont {Sch\"utz},
  \citenamefont {Genes},\ and\ \citenamefont
  {Pupillo}}]{hagenmuller2017cavity}%
  \BibitemOpen
  \bibfield  {author} {\bibinfo {author} {\bibfnamefont {D.}~\bibnamefont
  {Hagenm\"uller}}, \bibinfo {author} {\bibfnamefont {J.}~\bibnamefont
  {Schachenmayer}}, \bibinfo {author} {\bibfnamefont {S.}~\bibnamefont
  {Sch\"utz}}, \bibinfo {author} {\bibfnamefont {C.}~\bibnamefont {Genes}},\
  and\ \bibinfo {author} {\bibfnamefont {G.}~\bibnamefont {Pupillo}},\
  }\bibfield  {title} {\bibinfo {title} {Cavity-enhanced transport of charge},\
  }\href {https://doi.org/10.1103/PhysRevLett.119.223601} {\bibfield  {journal}
  {\bibinfo  {journal} {Phys. Rev. Lett.}\ }\textbf {\bibinfo {volume} {119}},\
  \bibinfo {pages} {223601} (\bibinfo {year} {2017})}\BibitemShut {NoStop}%
\bibitem [{\citenamefont {Semenov}\ and\ \citenamefont
  {Nitzan}(2019)}]{Semenov2019}%
  \BibitemOpen
  \bibfield  {author} {\bibinfo {author} {\bibfnamefont {A.}~\bibnamefont
  {Semenov}}\ and\ \bibinfo {author} {\bibfnamefont {A.}~\bibnamefont
  {Nitzan}},\ }\bibfield  {title} {\bibinfo {title} {Electron transfer in
  confined electromagnetic fields},\ }\href {https://doi.org/10.1063/1.5095940}
  {\bibfield  {journal} {\bibinfo  {journal} {The Journal of Chemical Physics}\
  }\textbf {\bibinfo {volume} {150}},\ \bibinfo {pages} {174122} (\bibinfo
  {year} {2019})}\BibitemShut {NoStop}%
\bibitem [{\citenamefont {Reitz}\ and\ \citenamefont
  {Genes}(2020)}]{10.1063/5.0033382}%
  \BibitemOpen
  \bibfield  {author} {\bibinfo {author} {\bibfnamefont {M.}~\bibnamefont
  {Reitz}}\ and\ \bibinfo {author} {\bibfnamefont {C.}~\bibnamefont {Genes}},\
  }\bibfield  {title} {\bibinfo {title} {Floquet engineering of molecular
  dynamics via infrared coupling},\ }\href {https://doi.org/10.1063/5.0033382}
  {\bibfield  {journal} {\bibinfo  {journal} {The Journal of Chemical Physics}\
  }\textbf {\bibinfo {volume} {153}},\ \bibinfo {pages} {234305} (\bibinfo
  {year} {2020})}\BibitemShut {NoStop}%
\bibitem [{\citenamefont {Kumar}\ \emph {et~al.}(2024)\citenamefont {Kumar},
  \citenamefont {Biswas}, \citenamefont {Rashid}, \citenamefont {Mony},
  \citenamefont {Chandrasekharan}, \citenamefont {Mattiotti}, \citenamefont
  {Vergauwe}, \citenamefont {Hagenmuller}, \citenamefont {Kaliginedi},\ and\
  \citenamefont {Thomas}}]{kumar2024extraordinary}%
  \BibitemOpen
  \bibfield  {author} {\bibinfo {author} {\bibfnamefont {S.}~\bibnamefont
  {Kumar}}, \bibinfo {author} {\bibfnamefont {S.}~\bibnamefont {Biswas}},
  \bibinfo {author} {\bibfnamefont {U.}~\bibnamefont {Rashid}}, \bibinfo
  {author} {\bibfnamefont {K.~S.}\ \bibnamefont {Mony}}, \bibinfo {author}
  {\bibfnamefont {G.}~\bibnamefont {Chandrasekharan}}, \bibinfo {author}
  {\bibfnamefont {F.}~\bibnamefont {Mattiotti}}, \bibinfo {author}
  {\bibfnamefont {R.~M.~A.}\ \bibnamefont {Vergauwe}}, \bibinfo {author}
  {\bibfnamefont {D.}~\bibnamefont {Hagenmuller}}, \bibinfo {author}
  {\bibfnamefont {V.}~\bibnamefont {Kaliginedi}},\ and\ \bibinfo {author}
  {\bibfnamefont {A.}~\bibnamefont {Thomas}},\ }\bibfield  {title} {\bibinfo
  {title} {Extraordinary electrical conductance through amorphous nonconducting
  polymers under vibrational strong coupling},\ }\href
  {https://doi.org/10.1021/jacs.4c03016} {\bibfield  {journal} {\bibinfo
  {journal} {Journal of the American Chemical Society}\ }\textbf {\bibinfo
  {volume} {146}},\ \bibinfo {pages} {18999} (\bibinfo {year} {2024})},\
  \bibinfo {note} {pMID: 38736166},\ \Eprint
  {https://arxiv.org/abs/https://doi.org/10.1021/jacs.4c03016}
  {https://doi.org/10.1021/jacs.4c03016} \BibitemShut {NoStop}%
\bibitem [{\citenamefont {Herrera}\ and\ \citenamefont
  {Owrutsky}(2020)}]{herrera2020molecular}%
  \BibitemOpen
  \bibfield  {author} {\bibinfo {author} {\bibfnamefont {F.}~\bibnamefont
  {Herrera}}\ and\ \bibinfo {author} {\bibfnamefont {J.}~\bibnamefont
  {Owrutsky}},\ }\bibfield  {title} {\bibinfo {title} {{Molecular polaritons
  for controlling chemistry with quantum optics}},\ }\href
  {https://doi.org/10.1063/1.5136320} {\bibfield  {journal} {\bibinfo
  {journal} {The Journal of Chemical Physics}\ }\textbf {\bibinfo {volume}
  {152}},\ \bibinfo {pages} {100902} (\bibinfo {year} {2020})},\ \Eprint
  {https://arxiv.org/abs/https://pubs.aip.org/aip/jcp/article-pdf/doi/10.1063/1.5136320/15572529/100902\_1\_online.pdf}
  {https://pubs.aip.org/aip/jcp/article-pdf/doi/10.1063/1.5136320/15572529/100902\_1\_online.pdf}
  \BibitemShut {NoStop}%
\end{thebibliography}%

\newpage
\onecolumngrid
\appendix
\section{Lindblad Derivation of Electron and Photon Currents}
\label{sec:currents}

Consider a nanojunctions where the system is charge coupled to macroscopic leads. Let us take the system and one of the leads as an open-quantum system described by the Hamiltonian $\hat{\mathcal{H}}=\hat{\mathcal{H}}_S+\hat{\mathcal{H}}_l+\hat{\mathcal{H}}_I
$. The system is described by the Hamiltonian $\hat{\mathcal{H}}_S$ contructed in terms of the Fermionic annihilation operator $\hat{c}_{i,\alpha}$ that annihilates electrons in the ground ($\alpha=g$) or the excited ($\alpha = e$) orbital of the $i$-th site of an array. The lead, either the left ($l=L$) or the right ($l=R$) lead, is described by the Hamiltonian $\hat{\mathcal{H}}_l=\sum_{k_l} \varepsilon_{k_l} \hat{c}_{k_l}^\dagger\hat{c}_{k_l}$, where $\hat{c}_{k_l}$ is the fermionic annihilation operator of electrons with energy $\varepsilon_{k_l}$. The interaction between the system and the lead is described by the interaction Hamiltonian
\begin{equation}\label{eq:interaction}
\hat{\mathcal{H}}_{I}=\sum_{\alpha, k_l} V_{k_l}\left(   \hat{c}_{j,\alpha}^\dagger  \hat{c}_{k_l} + \hat{c}_{j,\alpha}  \hat{c}_{k_l}^\dagger  \right) \equiv \mathcal{S} \otimes \mathcal{B} + \mathcal{S}^\dagger \otimes \mathcal{B}^\dagger.
\end{equation}
Eq. (\ref{eq:interaction}) describes the electron tunneling between the lead and all the orbital of the nearest $j$-th array site whit a rate $V_{k_l}$. Depending on the lead, the nearest site is $j=1$ for the left lead or $j=N$ for the right lead. Eq. (\ref{eq:interaction}) can be written in terms of system and bath operators, $\hat{\mathcal{S}}=\sum_\alpha \hat{c}_{j,\alpha}^\dagger$ and $\hat{\mathcal{B}}=\sum_{k_l} V_{k_l} \hat{c}_{k_l}$,
respectively.\\

In the Born approximation, system and leads are described by independent  density operators, $\hat{\rho}_S$ and 
\begin{equation}\label{eq:bath_state}
\hat{\rho}_B=\frac{e^{-( \hat{\mathcal{H}}_l-\mu_l \hat{\mathcal{N}}_l  )/T_0}}{{\rm Tr} \Big\{e^{-( \hat{\mathcal{H}}_l-\mu_l \hat{\mathcal{N}}_l  )/T_0} \Big\} },
\end{equation}
respectively. Leads are thermal ensemble of electrons at temperature $T_0$ and chemical potential $\mu_l$, written in terms of lead Hamiltonian $\hat{\mathcal{H}}_l$ and number operator $\hat{\mathcal{N}}_l = \sum_{k_l} \hat{c}_{k_l}^\dagger \hat{c}_{k_l}$. For weak-system bath coupling, the Markovian evolution of the system is given by the Redfield equation. The microscopic derivation of the system dynamics is well described in Ref.\cite{breuer2002theory}, recovering  the Lindblad master equation in Eq. (\ref{eq:QMeq}) when secular approximation is applied.\\
Here we derive the evolution of the mean-number of electrons in the lead $\langle \hat{\mathcal{N}}_l \rangle = {\rm Tr} \{ \hat{\mathcal{N}}_l \hat{\rho}_S \otimes \hat{\rho}_B \}$ , whit ${\rm Tr} \{ \}$ the trace, following the same microscopic derivation for the system dynamics. Therefore, using Redfield approach, the dynamics of $\langle \hat{\mathcal{N}}_l \rangle$ is given by \cite{Zeng-Zhao2019,PhysRevB.83.115414}
\begin{equation}\label{eq:def_number_evolution}
\frac{\rm d}{\rm dt}\langle \hat{\mathcal{N}}_l \rangle = - \int_{0}^\infty d\tau \hspace{0.1cm} {\rm Tr} \Big\{ \hat{\mathcal{N}}_l \left[ \widetilde{\mathcal{H}}_{I},\left[ \widetilde{\mathcal{H}}_{I}(-\tau),\hat{\rho}_S \otimes \hat{\rho}_B \right]  \right] \Big\},
\end{equation}
where $\widetilde{\mathcal{H}}_{I}(t)=\hat{\mathcal{U}}_t\hat{\mathcal{H}}_{I} \hat{\mathcal{U}}_t^\dagger \equiv \widetilde{\mathcal{S}}(t) \otimes \widetilde{\mathcal{B}}(t) + \widetilde{\mathcal{S}}^\dagger(t) \otimes \widetilde{\mathcal{B}}^\dagger(t)$ is the interaction Hamiltonian in the interaction picture obtained by the application of the unitary transformation $\hat{\mathcal{U}}_t=\exp (i[ \hat{\mathcal{H}}_S +  \hat{\mathcal{H}}_B] t)$. The associated time-dependent system and bath operators are $\widetilde{\mathcal{S}}(t) = \hat{\mathcal{U}}_t\hat{\mathcal{S}}\hat{\mathcal{U}}_t^\dagger $ and  $\widetilde{\mathcal{B}}(t) = \hat{\mathcal{U}}_t\hat{\mathcal{B}}\hat{\mathcal{U}}_t^\dagger $, respectively.

By inserting interaction Hamiltonian in terms of the system and the bath operators, Eq. (\ref{eq:def_number_evolution}) is reduced to
\begin{equation}\label{eq:number2}
\begin{split}
\frac{{\rm d}}{\rm dt}\langle \hat{\mathcal{N}}_l\rangle=-\int_{0}^\infty d\tau\hspace{0.1cm}\Big[& \langle \hat{\mathcal{S}} \widetilde{\mathcal{S}}^\dagger (-\tau) \rangle \langle \hat{\mathcal{N}}_l \hat{\mathcal{B}} \widetilde{\mathcal{B}}^\dagger (-\tau)\rangle +\langle \hat{\mathcal{S}}^\dagger \widetilde{\mathcal{S}}(-\tau) \rangle\langle \hat{\mathcal{N}}_l \hat{\mathcal{B}}^\dagger \widetilde{\mathcal{B}} (-\tau)\rangle\\
-& \langle \widetilde{\mathcal{S}}^\dagger (-\tau)\hat{\mathcal{S}}\rangle
\langle \widetilde{\mathcal{B}}^\dagger(-\tau)\hat{\mathcal{N}}_l \hat{\mathcal{B}} \rangle- \langle\widetilde{\mathcal{S}} (-\tau) \hat{\mathcal{S}}^\dagger \rangle \langle \widetilde{\mathcal{B}}(-\tau)\hat{\mathcal{N}}_l \hat{\mathcal{B}}^\dagger \rangle \\
-& \langle \hat{\mathcal{S}}^\dagger \widetilde{\mathcal{S}}(-\tau)\rangle\langle \hat{\mathcal{B}}^\dagger \hat{\mathcal{N}}_l \widetilde{\mathcal{B}}(-\tau) \rangle-\langle \hat{\mathcal{S}} \widetilde{\mathcal{S}}^\dagger(-\tau)\rangle\langle \hat{\mathcal{B}}\hat{\mathcal{N}}_l \widetilde{\mathcal{B}}^\dagger(-\tau) \rangle\\
+& \langle \widetilde{\mathcal{S}}(-\tau) \hat{\mathcal{S}}^\dagger \rangle\langle \widetilde{\mathcal{B}} (-\tau) \hat{\mathcal{B}}^\dagger \hat{\mathcal{N}}_l \rangle + \langle \widetilde{\mathcal{S}}^\dagger(-\tau) \hat{\mathcal{S}} \rangle \langle \widetilde{\mathcal{B}}^\dagger (-\tau) \hat{\mathcal{B}}\hat{\mathcal{N}}_l \rangle \Big].
\end{split}
\end{equation}
Using that $\tilde{\mathcal{B}}(t) =\sum_{k_l}V_{k_l}\exp (-i\omega_{k_l}t)\hat{c}_{k_l}$, the bath terms in Eq. (\ref{eq:number2}) reduce it to
\begin{equation}\label{eq:number3}
\begin{split}
\frac{{\rm d}}{\rm dt}\langle \hat{\mathcal{N}}_l\rangle=-\int_{0}^\infty d\tau\hspace{0.1cm}\Big[&\langle \hat{\mathcal{S}}^\dagger \widetilde{\mathcal{S}}(-\tau) \rangle \mathcal{C}(\tau)-\langle\widetilde{\mathcal{S}} (-\tau) \hat{\mathcal{S}}^\dagger \rangle \overline{\mathcal{C}}(-\tau)-\langle \hat{\mathcal{S}} \widetilde{\mathcal{S}}^\dagger(-\tau)\rangle \overline{\mathcal{C}}(\tau)+\langle \widetilde{\mathcal{S}}^\dagger(-\tau) \hat{\mathcal{S}} \rangle \mathcal{C}(-\tau) \Big],
\end{split}
\end{equation}
where the thermal structure of the baths in Eq. (\ref{eq:bath_state}) leads to the bath correlations
\begin{eqnarray}\label{eq:correlations}
\mathcal{C}(\tau) &=& \sum_{k_l} |V_{k_l}|^2 f_l(\varepsilon_{k_l}) \exp\left( i \varepsilon_{k_l} \tau\right),\\
\overline{ \mathcal{C}}(\tau) &=& \sum_{k_l} |V_{k_l}|^2 [ 1-f_l(\varepsilon_{k_l}) ]\exp\left(-i \varepsilon_{k_l} \tau\right),
\end{eqnarray}
written in terms of the Fermi distribution of the lead $f_l(\omega )=\left[  \exp{((\omega-\mu_l)/T_0)}+1 \right]^{-1}$.

We use the system eigenbasis, i.e., $\hat{\mathcal{H}}_S= \sum_\epsilon \omega_\epsilon \ket{\epsilon}\bra{\epsilon}$, to write the system operators as $ \hat{\mathcal{S}}=\sum_{\epsilon,\epsilon'} \bra{\epsilon}\sum_\alpha \hat{c}_{j,\alpha}^\dagger \ket{\epsilon'}  \ket{\epsilon}\bra{\epsilon'}$ and $\widetilde{\mathcal{S}}(t) =\hat{\mathcal{U}}_t \hat{\mathcal{S}}\hat{\mathcal{U}}_t^\dagger =\sum_{\epsilon,\epsilon'} \bra{\epsilon}\sum_\alpha \hat{c}_{i,\alpha}^\dagger \ket{\epsilon'} \exp (i\omega_{\epsilon,\epsilon'}t)\ket{\epsilon}\bra{\epsilon'}$, where $\omega_{\epsilon,\epsilon'} = \omega_\epsilon-\omega_{\epsilon'}$ is a transition frequency between eigenstates. It reduces the system terms in Eq. (\ref{eq:number3}) to
\begin{eqnarray}
\langle \hat{\mathcal{S}}^\dagger (-\tau)\hat{\mathcal{S}}\rangle &=& \langle \hat{\mathcal{S}}^\dagger\hat{\mathcal{S}}(-\tau)\rangle^* = \sum_{\epsilon,\epsilon'}|\bra{\epsilon}\sum_\alpha \hat{c}_{i,\alpha}^\dagger \ket{\epsilon'} |^2\exp(i\omega_{\epsilon,\epsilon'}\tau)\rho_{\epsilon',\epsilon'}, \label{eq:system_1}\\
\langle \hat{\mathcal{S}} (-\tau)\hat{\mathcal{S}}^\dagger\rangle &=& \langle \hat{\mathcal{S}}\hat{\mathcal{S}}^\dagger(-\tau)\rangle^* = \sum_{\epsilon,\epsilon'}|\bra{\epsilon}\sum_\alpha \hat{c}_{i,\alpha}^\dagger \ket{\epsilon'} |^2\exp(-i\omega_{\epsilon,\epsilon'}\tau)\rho_{\epsilon,\epsilon}.\label{eq:system_2}
\end{eqnarray}
The last expressions are just written in terms of eigenstate population $\rho_{\epsilon,\epsilon}\equiv \langle \epsilon | \hat{\rho}_S \ket{\epsilon}$. Any contribution from the eigenstate coherence $\rho_{\epsilon,\epsilon'}\equiv \langle \epsilon | \hat{\rho}_S \ket{\epsilon'}$ has been ignored because secular approximation.\\

The final step uses the principal value theorem
\begin{equation}
\int_0^\infty d\tau e^{\pm i \omega \tau} = \pi \delta (\omega) \pm i{\rm P.V.} \frac{1}{\omega},
\end{equation}
where the $\delta (\omega)$ is the Dirac delta and ${\rm P.V.}$ is the principal value. Applying the last formula, the integrals over the bath correlations in Eq. (\ref{eq:number3}),
\begin{eqnarray}
\int_0^\infty d\tau \hspace{0.1cm} \mathcal{C}(\pm \tau)\exp\left(\mp i\omega_{\epsilon,\epsilon'}\tau\right) &=&\frac{1}{2}\Gamma_l f_l(\omega_{\epsilon,\epsilon'}),\label{eq:rate_1}\\
\int_0^\infty d\tau  \hspace{0.1cm} \overline{\mathcal{C}} (\pm \tau)\exp\left(\pm i\omega_{\epsilon,\epsilon'}\tau\right) &=&\frac{1}{2}\Gamma_l \left[1-f_l(\omega_{\epsilon,\epsilon'})\right],\label{eq:rate_2}
\end{eqnarray}
can be written in terms of the lead tunneling rate $\Gamma_l(\omega)=\sum_{k_l} |V_{k_l}|^2 \delta(\omega-\omega_{k_l}) \equiv \Gamma_l$, which is assumed to be constant (wideband approximation). Imaginary terms are ignored as we did in the system dynamics because they leads to negligible Lamb shifts of the eigenstate frequencies.

Using system terms in Eq. (\ref{eq:system_1},\ref{eq:system_2}) and bath terms in Eq. (\ref{eq:rate_1},\ref{eq:rate_2}), the dynamics of electron number in the leads, given by Eq. (\ref{eq:number3}), is reduced to
\begin{equation}\label{eq:dynamics_bath_number}
\frac{\rm d}{\rm dt}\langle \hat{\mathcal{N}}_l \rangle=\Gamma_l\sum_{\epsilon,\epsilon'}|\bra{\epsilon}( \hat{c}^\dagger_{i,g}+\hat{c}^\dagger_{i,e})\ket{\epsilon'}|^2\Big\{ \left[ 1-f_l(\omega_{\epsilon,\epsilon'})\right] \rho_{\epsilon,\epsilon} - f_l(\omega_{\epsilon,\epsilon'}) \rho_{\epsilon',\epsilon'} \Big\}.
\end{equation}
Equation (\ref{eq:dynamics_bath_number}) recovers expressions for left and right current, $I_L = - {\rm d} \langle \hat{\mathcal{N}}_L \rangle/ {\rm dt} $ and $I_R = {\rm d} \langle \hat{\mathcal{N}}_R \rangle/ {\rm dt}$, respectively, in the main text. For deriving $j_r =  {\rm d} \langle \hat{\mathcal{N}}_{\rm r} \rangle/ {\rm dt} $, shown in Eq. (\ref{eq:luminescence projected}), we use the same procedure described above, starting from the evolution of the photon number in the EM field $\hat{\mathcal{N}}_{\rm r} = \sum_p \hat{a}_p^\dagger \hat{a}_p$, with bosonic operators $\hat{a}_p$ of the photonic mode $\omega_p$.

\end{document}